\documentclass[prb,amsmath,superscriptaddress,twocolumn]{revtex4}
\usepackage{amsmath}
\usepackage{bm}
\usepackage{colordvi}
\usepackage{color}
\usepackage{amssymb}
\usepackage{graphicx}
\usepackage{epstopdf}

\newcommand{\tB}{\widetilde B}

\newcommand{\tq}{\widetilde q}
\newcommand{\tQ}{\widetilde {\cal Q}}

\newcommand{\bl}{\bar l}
\newcommand{\br}{\bar r}
\newcommand{\bb}{\bar b}

\newcommand{\bbk}{\bar k}

\newcommand{\bxi}{\bar\xi}
\newcommand{\lr}{\langle}
\newcommand{\rr}{\rangle}
\newcommand{\cT}{{\cal T}}
\newcommand{\oB}{{\overline B}}
\newcommand{\ob}{{\overline b}}
\newcommand{\oq}{{\overline q}}
\newcommand{\oQ}{{\overline {\cal Q}}}
\newcommand{\oT}{{\overline T}}
\begin{document}

\title{Generalized Landauer formula for time-dependent potentials and noise-induced zero-bias dc current}

\author{Shmuel Gurvitz}
\email{shmuel.gurvitz@weizmann.ac.il}

\affiliation{Department of Particle Physics and Astrophysics,\\  Weizmann Institute of
Science, Rehovot 76100, Israel}
\affiliation{ Beijing Computational Science Research Center, Beijing 100084, China}
\date{\today}

\textbf{}

\begin{abstract}
Using a new developed Single-Electron approach, we derive the Landauer-type formula for electron transport in arbitrary time-dependent potentials. This formula is applied for randomly fluctuating potentials represented by a dichotomic noise. We found that the noise can produce dc-current in quantum system under zero-bias voltage by breaking the time-reversal symmetry of the transmission coefficient. We show that this effect is due to  decoherence, produced by the noise, which can take place in many different systems.
\end{abstract}

\maketitle
\section{Introduction}
\label{sec1}
Directed particle flow in system at equilibrium (zero-bias voltage), induced by an external periodical force, represents one of the most interesting and important effects in non-equilibrium quantum transport \cite{brouwer,hanggi1,cubero}. The same effect, but induced by the isotropic {\em randomly fluctuating} environment could be even more important, especially for biological systems \cite{suarez,rafa1}.

These and similar phenomena can be investigated by electron transport in mesoscopic  systems under external time-dependent field. For that study one can use different approaches, like non-equilibrium Green's function (NEGF) \cite{1}, hybrid Floquet-NEGF treatments \cite{5,6}, time-dependent scattering state methods\cite{tu} and others. Most of them are quite complicated for applications. For this reason, a more convenient, Markovian Master equations approach, has been widely used. However, this approach is valid for large bias limit \cite{gur}, namely for $\Gamma/V\ll 1$ where $\Gamma$ is the energy levels width and $V$ is the bias-voltage. Therefore it would not be suitable for zero-bias case.

In contrast, the Landauer approach to non-interacting electron transport \cite{imry} is free from this restriction. Indeed, the celebrated Landauer formula, relating the steady-state current to quantum transmission coefficients, is valid for any bias. However, its generalization to driven quantum transport is not straightforward. For instance, it cannot be performed  without taking into account the many-particle effects \cite{hanggi}. Such an extension has been done for periodically driven quantum systems in a framework of the Floquet theory, by using the NEGF technique \cite{hanggi2,hanggi}. As a result, the static transmission coefficients are replaced by the time-dependent transmissions in the modified Landauer formula. However, such a treatment cannot be easily adapted for arbitrary time-dependent drive, in particular for the case of randomly fluctuated energy levels and tunneling barriers.

In the present paper we use a recently proposed single-electron  approach (SEA)  \cite{single}. This approach has been derived directly from the time-dependent Schr\"odinger equation by using the single-electron {\em Ansatz} for the many-electron wave function. The SEA is very different from standard NEGF techniques,  and it does not involve the Floquet expansion. As a result, it yields a new generalized Landauer formula for the transient current in {\em any} time-dependent potentials. This formula is equally suitable for study electron transport in periodically modulated or in randomly fluctuating potentials \cite{single,GAE}.

In this paper we extend our previous result to a general, non-Markovian environment. Then we concentrate on the zero-bias dc-current through single and double-dot systems, driven by an external dichotomic (telegraph) noise. A different behaviour of the both systems displays quantum nature of the zero-bias current. We show that this is a generic quantum-mechanical effect, associated with decoherence.

The plan of this paper is as following. Sec.~\ref{sec2} deals with an electron motion through quantum dot, coupled to two leads with arbitrary spectral densities. The single-electron wave-function is obtained by solving time-dependent Schr\"odinger equation with the time-dependent Hamiltonian.

In Sec.~\ref{sec3} we introduce pure and mixed (finite temperature) many-particle states of an entire system and evaluate the time-dependent charges and currents from the many-particle wave-function. Sec.~\ref{sec4} deals with a single quantum dot under the telegraph noise and the time-dependent ensemble averaged current, flowing through this system.

Section~\ref{sec5} considers the electric current through a double-dot under the telegraph noise, by concentrating on the zero-bias current in the steady-state limit. Simple analytical expression is obtained for this quantity, which is compared with the exact numerical solution. Sec.~\ref{sec6} discusses generic quantum-mechanical mechanism of the zero-bias current, induced by the noise. The results and their experimental meaning are discussed in Secs.~\ref{sec4} and \ref{sec7}.

Derivation of the Shapiro-Loginov differential formula for a finite temperature noise is presented in Appendix~\ref{app1}. Appendix~\ref{app2} contains some details of derivation of the single-electron Master equations used in our treatment.

\section{Single-electron motion through quantum dot}
\label{sec2}

\begin{figure}[h]
\includegraphics[width=7cm]{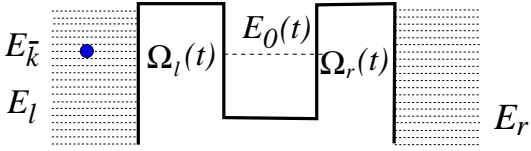}
\caption{(Color online) Quantum dot coupled with two leads, where $E_{l(r)}$ denote the energy levels in the left (right) lead. A single electron occupies the energy level $E_{\bbk}$ in the left lead at $t=0$.}
\label{fg1}
\end{figure}

Consider a single quantum dot coupled with two reservoirs, Fig.~\ref{fg1},
and described by the following Hamiltonian
\begin{align}
&H(t)=\sum_lE_l\hat c_l^\dagger\hat c_l+\sum_rE_r\hat c_r^\dagger \hat
c_r+E_0(t)\hat c_0^\dagger \hat c_0
\nonumber\\
&+\left(\sum_{l}\Omega_l(t)\hat
c^\dagger_l \hat c_0 +\sum_{r}\Omega_r(t)\hat c^\dagger_r \hat c_0+H.c.\right)\, .
\label{ham}
\end{align}
Here $\hat c_{l(r)}$ denotes the electron annihilation operator in the left (right) lead and $\hat c_{0}$ is the same for the quantum dot. The tunneling coupling of the left (right) lead with the dot, $\Omega^{}_{l(r)}(t)$, are real valued. These couplings and the energy level of the dot, $E_0(t)$, are time-dependent, where the energy levels of the leads, $E_{l,r}$, are time-independent.

The time-dependent Hamiltonian~(\ref{ham}) can be realized experimentally via time-dependent gate voltage applied to each of the barriers and to the dot. Note that when the gate voltage is applied only to the dot, $E_0\to E_0^{}(t)$, it generates time-dependence of the tunneling coupling $\Omega_{l,r}^{}$, as well. It can be seen from the Bardeed formula \cite{bardeen,gur1}. Indeed, in the semiclassical limit it gives $\Omega_{l,r}^{}\propto\exp \big(-\kappa_{L,R}^{}{\cal L}_{L,R}^{}\big)$, where ${\cal L}_{L,R}^{}$ is the barrier width, $\kappa_{L,R}^{}=\sqrt{2m[v_{L,R}^{}-E_0^{}(t)]}$ and   $v_{L,R}^{}$ is the barrier hight. However, in the limit $\delta E_0^{}/(v_{L,R}^{}-E_0^{}]\ll 1$, where $\delta E_0^{}$, is a variation of the energy level with time, the induced time-dependence of tunneling couplings can be neglected.

Similarly, when the voltage is applied to the barrier only, $v_{L,R}\to v_{L,R}(t)$, it generates time-dependence of the dot's energy level. It can be seen from matching of logarithmic derivative at the dot's boundary. However, if $\delta v_{L,R}/(v_{L,R}-E_0]\ll 1$ (high barrier), the induced time-dependence of the energy level $E_0$ can be disregarded. All this implies that for high barriers, the energy dependence of tunneling couplings remains the same as for time-independent barriers. As a result, one can write
\begin{align}
\Omega_{r,l}^{}(t)\equiv\Omega_{l,r}^{}\, w_{L,R}^{}(t)\, ,
\label{omtp}
\end{align}
where $w_{L,R}^{}(t)$ accounts variation of the barrier hight with time.

\subsection{Single-electron wave function.}
\label{sec2a}

It is demonstrated in Refs.~[\onlinecite{single,GAE}] that the time-dependent electron current of  non-interacting electrons, flowing through the quantum dot, Fig.~(\ref{fg1}), is totally determined by a single-electron wave function, $|\psi^{(\bbk)}_{}(t)\rangle$. The latter is obtained from the time-dependent Schr\"odinger equation,
\begin{align}
i\,{d\over dt}|\psi_{}^{(\bbk)}(t)\rangle =H(t)|\psi_{}^{(\bbk)}(t)\rangle\, ,
\label{sel}
\end{align}
where the index $\bbk=\{\bl,\br,0\}$ denotes the electron's  initial state, corresponding to the occupied level $E_{\bl(\br)}$ in the left (right) lead,  or the level $E_{0}$ in the quantum dot, Fig.~\ref{fg1}, at $t=0$.

In order to solve Eq.~(\ref{sel}), we represent the single-electron wave function, $|\psi^{(\bbk)}_{}(t)\rr$, in the basis of the Hamiltonian (\ref{ham}),
\begin{align}
&|\psi_{}^{(\bbk)}(t)\rr=\hat\Phi_{}^{(\bbk)\dagger}(t)|0\rr~~~{\rm with}\nonumber\\
&\hat\Phi_{}^{(\bbk)\dagger}(t)=\sum_l b_l^{(\bbk)}(t)c_l^\dagger+b_0^{(\bbk)}(t)c_0^\dagger+\sum_r b_r^{(\bbk)}(t)c_r^\dagger
\label{wf1}
\end{align}
where $b_{l(r)}^{(\bbk)}(t)$ and $b_{0}^{(\bbk)}(t)$ are  probability amplitudes of finding the electron in the left (right) lead at the level $E_{l(r)}$, or inside the dot at the level $E_0$, respectively, for the initial conditions
\begin{align}
b_{l}^{(\bbk)}(0)=\delta^{}_{\bbk l}\delta^{}_{{\bar l}l},\ \  b_{r}^{(\bbk)}(0)=\delta^{}_{\bbk r}\delta^{}_{{\bar r}r},\ \   b_{0}^{(\bbk)}(0)=\delta^{}_{\bbk 0}.
\label{incond0}
\end{align}

Substituting Eq.~(\ref{wf1}) into Eq.~(\ref{sel}) we obtain the following set of coupled linear differential equations for amplitudes $b(t)$,
\begin{subequations}
\label{c6}
\begin{align}
i\,\dot {b}_{l}^{(\bbk)}(t)&=E^{}_lb_{l}^{(\bbk)}(t)
+\Omega^{}_l(t)\,b_{0}^{(\bbk)}(t)\ ,
\label{c6a}\\
i\,\dot {b}_{0}^{(\bbk)}(t)&=E^{}_0(t)\,b_{0}^{(\bbk)}(t)\nonumber\\
&
+\sum_l\Omega^{}_l(t)\,b_{l}^{(\bbk)}(t)
%\nonumber\\
+\sum_r\Omega^{}_r(t)\,b_{r}^{(\bbk)}(t)\ ,
\label{c6b}\\
i\,\dot {b}_{r}^{(\bbk)}(t)&=E^{}_r\,b_{r}^{(\bbk)}(t)
+\Omega^{}_r(t)\,b_{0}^{(\bbk)}(t)\ .\label{c6c}
\end{align}
\end{subequations}

Equations~(\ref{c6a}) and (\ref{c6c}) for the amplitudes $b_{p}^{(\bbk)}(t)$, where $p=\{l,r\}$, can be solved explicitly, thus obtaining
\begin{align}
b_{p}^{(\bbk)}(t)=e^{-iE_{p}t}\Big[\delta^{}_{p\bbk}
-\int\limits_{0}^t
i\,\Omega_{p}(t') b_{0}^{(\bbk)}(t')e^{iE_{p}t'}dt'\Big]
\label{c12p}
\end{align}
It is useful to represent the amplitude $b_{0}^{(\bbk)}(t)$ by
\begin{align}
b_{0}^{(\bbk)}(t)=b_0^{(\alpha)}(E_{\bbk},t)e^{-iE_{\bbk}t}
\label{alb}
\end{align}
where $\alpha =L,R$ denotes the lead, occupied initially by electron at the level $E_{\bbk}$. Substituting Eq.~(\ref{c12p}) into Eq.~(\ref{c6b}) and using Eq.~(\ref{alb}) we find,
\begin{align}
&{d\over dt}\,b_0^{(\alpha)}(E_{\bbk},t)
=i\big[E_{\bbk}-E^{}_0(t)\big]b_0^{(\alpha)}(E_{\bbk},t)-
i\,\Omega^{}_{\bbk}(t)\nonumber\\
&-\int\limits_0^t\big[G_L(t,t')+G_R(t,t')\big]
e^{iE_{\bbk}(t-t')}\,b_0^{(\alpha)}(E_{\bbk},t')dt'\, ,
\label{gg}
\end{align}
with $\bbk\in\alpha$ and
\begin{align}
G_{\alpha'}(t,t')=\sum_{p\in\alpha'}\Omega_p(t)\Omega_p(t')
e^{iE_p(t'-t)}\, .
\end{align}
Here $\alpha'=L,R$. In the continuous limit, $\sum_{p\in\alpha'}\to\int\varrho_{\alpha'}(E_p)dE_p$, one can write
\begin{align}
G_{\alpha'}(t,t')= \int\limits_{-\infty}^\infty
\Omega_p(t)\Omega_p(t')e^{iE_p(t'-t)}
\varrho_{\alpha'}^{}(E_p)\,dE_p
\label{g}
\end{align}
where $\varrho_{\alpha'}(E_p)$ is the density of state of a lead $\alpha'$.

\subsection{Charges and currents.}
\label{sec2b}

Solving the integro-differential Eq.~(\ref{gg}), we obtain the amplitude $b^{(\alpha)}_{0}(E_{\bbk},t)$ and therefore the probability of finding the dot occupied,
\begin{align}
&q_0^{(\alpha)}(E_{\bbk},t)=\lr\psi_{\bbk}(t)|\hat c^\dagger_0\hat c_0^{}|\psi_{\bbk}(t)\rr=|b^{(\alpha)}_{0}(E_{\bbk},t)|^2\, .
\label{q01}
\end{align}
Using Eq.~(\ref{gg}) and (\ref{q01}) one can derive the following useful relation for time-derivative of this quantity
\begin{align}
&{d\over dt}\,q_0^{(\alpha)}(E_{\bbk},t)=-2{\rm Im}\big[b^{(\alpha)}_{0}(E_{\bbk},t)\big]
\Omega_{\alpha}(t)
\nonumber\\
&-2{\rm Re}
\int\limits_0^t b_0^{(\alpha)*}
(E_{\bbk},t)b_0^{(\alpha)}(E_{\bbk},t') G(t,t')
e^{iE_{\bbk}(t-t')}dt'
\label{dq0}
\end{align}
where $\bbk\in\alpha$ and $G(t,t')=G_L(t,t')+G_R(t,t')$.

Consider now the single-electron current in the lead $\alpha'=L,R$ (in units of the electron charge), given by
\begin{align}
&{\cal I}_{\alpha'} (E_{\bbk},t)= i\lr\psi_{\bbk}(t)\Big|\Big[H(t),\sum_{p\in\alpha'}\hat c^\dagger_p\hat c_p^{}\Big]\Big|\psi_{\bbk}(t)\rr\nonumber\\
&=i\sum_{p\in\alpha'}\Omega_p(t)\lr\psi_{\bbk}(t)
(\hat c^\dagger_p \hat c_0^{}-\hat c^\dagger_0 \hat c_p^{})|\psi_{\bbk}(t)\rr\nonumber\\
&=2\,{\rm Im}\sum_{p\in\alpha'}\Omega_p(t)
b_0^{(\bbk)}(t)b_p^{(\bbk)*}(t)
\label{cali1}
\end{align}
Note that in Eqs.~(\ref{q01})-(\ref{cali1}) the index $\alpha'$ denotes the lead where the current (${\cal I}_{\alpha'}$) is evaluated, whereas the indices $\alpha$ and $\bbk\in\alpha$ denote the lead and energy level ($E_{\bbk}$), occupied by electron at $t=0$. We also point out that the electron's initial state is {\em not} an eigen-state of the total Hamiltonian. Therefore the electron's wave function in the final state is spread over different energy levels of the lead, even for the time-independent Hamiltonian.

In addition we emphasize that the continuous limit corresponds to the leads size increasing to infinity. Otherwise the leads spectrum remains discrete, so that  the electron's wave function cannot reach the steady state limit at  $t\to\infty$. Thus, the steady-state limit always implies the following order of the limits: first, the size of leads goes to infinity and then $t\to\infty$.

Using Eqs.~(\ref{c12p})-(\ref{g}), we can rewrite the single-electron current as
\begin{align}
&{\cal I}_{\alpha'} (E_{\bbk},t)=2\,{\rm Im}\big[b_0^{(\alpha)}(E_{\bbk},t)\big]\Omega_{\bbk}(t)
\delta_{\alpha\alpha'}\nonumber\\
&+2\,{\rm Re}\int\limits_0^t b_0^{(\alpha)*}(E_{\bbk},t)\, b_0^{(\alpha)}(E_{\bbk},t')G_{\alpha'}^{}(t,t')
e^{iE_{\bbk}(t-t')}dt'\, ,
\label{secur}
\end{align}
where ${\bbk}\in\alpha$. Hence, for $\alpha'\not =\alpha$, the current ${\cal I}_{\alpha'} (E_{\bbk},t)$ is given by the last term of Eq.~(\ref{secur}) only. Such a  single-electron  current represents time-dependent extension of the quantum-mechanical transmission probability (transmission coefficient), defined as
\begin{subequations}
\label{transm1}
\begin{align}
T_{L\to R}(E_{\bar l},t)&=2\pi\varrho_L^{}(E_{\bl})\,
{\cal I}_R(E_{\bl},t),\label{transm1a}\\
T_{R\to L}(E_{\bar r},t)&=2\pi\varrho_R^{}(E_{\br})\,
{\cal I}_L(E_{\br},t)
\label{transm1b}
\end{align}
\end{subequations}
The first one, $T_{L\to R}$, is transmission probability from the level $E_{\bl}^{}$ in the left lead to the right lead. Respectively, the second one, $T_{R\to L}$, is transmission probability from the level $E_{\br}^{}$ in the right lead to the left lead. It is demonstrated below that for {\em time-independent} Hamiltonian,
\begin{align}
T_{L\to R}(E,t\to\infty)=T_{R\to L}(E,t\to\infty)=\bar T(E)\, ,
\label{timeind}
\end{align}
where $\bar T(E)$ is the standard quantum mechanical transmission coefficient. Note that Eq.~(\ref{timeind}) displays the time-reversal symmetry of the transmission probability,, valid for any time-independent Hamiltonian.

Let us consider the single-electron currents, ${\cal I}_L(E_{\bl},t)$ and  ${\cal I}_R(E_{\br},t)$, when the initial state belongs to the same lead,  $\alpha=\alpha'$, Eq.~(\ref{secur}). In this case, the current is given by two terms of Eq.~(\ref{secur}). Using Eq.~(\ref{dq0}) we obtain
\begin{align}
{\cal I}_R(E_{\br},t)=-{d\over dt}\,q_0^{(R)}(E_{\br},t)-{\cal I}_L(E_{\br},t)\, .
\label{opt}
\end{align}
The same expression is found for ${\cal I}_L(E_{\bl},t)$, by interchanging $R\leftrightarrow L$ and $\br\leftrightarrow \bl$.

\subsection{Wide-band limit.}
\label{sec2c}

In the Markovian case (wide-band limit), the spectral-density function, $\Omega_p^2(t)\varrho(E_p)$ is independent of energy \cite{fn0}. 
This implies that $\Omega_p^{}(t)\Omega_p^{}(t')\varrho_{\alpha'}^{}(E_p)
=\Omega_{\alpha'}^{}(t)\Omega_{\alpha'}^{}(t')\varrho_{\alpha'}^{}$ in  Eq.~(\ref{g}). As a result,
\begin{align}
G_{\alpha'}(t,t')=2\pi\Omega_{\alpha'}(t)\Omega_{\alpha'}(t')\varrho_{\alpha'} \delta(t'-t).
\label{gal}
\end{align}
Then using $\int_0^t b_0^{(\alpha)}(E_{\bbk},t')\delta(t'-t)dt'=b_0^{(\alpha)}
(E_{\bbk},t)/2$, we reduce Eq.~(\ref{gg}) to the following equation
\begin{align}
{d\over dt}b^{(\alpha)}_{0}(E_{\bbk},t)
&=i\Big[E_{\bbk}-E^{}_0(t)+i{\Gamma(t)\over2}\Big]
b_{0}^{(\alpha)}(E_{\bbk},t)\nonumber\\
&~~~~~~~~~~~~~~~~~~~~~~~~~~~~~~~~~
-i\Omega^{}_{\alpha}(t) ,
\label{gg1}
\end{align}
where
\begin{align}
\Gamma(t)\equiv \Gamma_L(t)+\Gamma_R(t)=2\pi\Omega_{L}^{2}(t)\varrho_{L}^{}
+2\pi\Omega_{R}^{2}(t)\varrho_{R}^{}\, ,
\label{gamtot}
\end{align}
is total time-dependent width of the level $E_0(t)$.
The time-dependent transmission coefficients, Eqs.~(\ref{transm1} a,b) are given by
\begin{align}
&T_{L\to R}(E_{\bar l},t)=2\pi\varrho_L^{}q_0^{(L)}(E_{\bl},t)\Gamma_R(t),\nonumber\\
&T_{R\to L}(E_{\bar r},t)=
2\pi\varrho_R^{}q_0^{(R)}(E_{\br},t)\Gamma_L(t)
\label{trans3}
\end{align}
where $q_0^{(\alpha)}(E_{\bbk},t)= |b_0^{(\alpha)}(E_{\bbk},t)|^2$ is probability of finding the dot occupied, Eq.~(\ref{q01}), by an electron coming from the lead $\alpha$.

Equation~(\ref{gg1}) can be solved straightforwardly. One finds \cite{single,GAE},
\begin{align}
q^{(\alpha)}_{0}(E_{\bbk},t)=\left |\int\limits_0^t\Omega_{\alpha}^{}(t')
e^{i[{\cal E}_0(t')-E_{\bbk}]t'-i{\cal E}_0(t)t}dt'\right|^2\, ,
\label{oc1}
\end{align}
where
\begin{align}
{\cal E}_0(t)\,t=\int\limits_0^t \left [E_0(t')-i{\Gamma (t')\over2}\right]dt'\, .
\label{calen}
\end{align}
If the electron is  initially inside the dot, $\bbk=0$, then
\begin{align}
q_0^{(0)}(t)=q_0^{(0)}(0)e^{-\int_0^t \Gamma(t')dt'}_{}\, .
\label{oc2}
\end{align}

In the case of time-independent Hamiltonian, one easily obtains from Eqs.~(\ref{oc1})-(\ref{calen}) that $T_{L\to R}=T_{R\to L}\equiv T(E,t)$, where
\begin{align}
T(E,t)={\Gamma_L\Gamma_R\over (E-E_0)^2+{\Gamma^2\over4}}\big[1-2\cos (Et)e^{-\Gamma t/2}+e^{-\Gamma t}\big]
\label{brwig}
\end{align}
In the asymptotic limit, $t\to\infty$, this expression reproduces a well-known Breit-Wigner formula for resonant transmission, $\bar T(E)=\Gamma_L\Gamma_R/\big[ (E-E_0)^2+{\Gamma^2\over4}\big]$.

\subsection{Finite-band leads.}
\label{sec2d}

Consider now the leads of a finite band-width $W_a^{}$, centered at $\epsilon_a^{}$, where $a=\{L,R\}$. For simplicity, we assume high barrier hight, so $\Omega_{l,r}^{}(t)$ are represented by Eq.~(\ref{omtp}), where $\Omega_{l,r}^{}\equiv\Omega_{L,R}^{}(E_{l,r}^{})$ is the tunneling coupling for a static barrier. The lead's spectra1 function is usually parameterized as
\begin{align}
\Omega_a^2(E_p)\varrho_a(E_p)={\Gamma_a\over 2\pi}\sqrt{1-{(E_p-\epsilon_a^{})^2\over W^2_a}}\, ,
\label{fb}
\end{align}
corresponding to a semi-infinite lead, consisted of periodic one-dimensional chain of quantum wells with the nearest-neighbor coupling.

Substituting Eqs.~(\ref{omtp}), (\ref{fb}) into Eq.~(\ref{g}) we can evaluate $G_a(t,t')$, and then insert it into the integro-differential equation~(\ref{gg}). Unfortunately, despite its simple form, the spectral density (\ref{fb}) does not allow us to solve Eq.~(\ref{gg}) analytically. For this reason, we approximate Eq.~(\ref{fb}) by a Lorentzian
\begin{align}
\Omega_a^2(E_p)\varrho_a(E_p)
={\Gamma_a\over2\pi}{\Lambda_a^2\over (E_p-\epsilon_a^{})^2+\Lambda_a^2}\, ,
\label{lor}
\end{align}
with $\Lambda_a=\sqrt{2}\,W_a$. The latter provides the same curvature at the band center, making the Lorentzian~(\ref{lor}) a very good approximation for a finite range spectral function~(\ref{fb}) (see Ref.~[\onlinecite{g1}]).

Now we demonstrate that the Lorentzian form of spectral-density function (\ref{lor}) allows us to reduce the integro-differential equation~(\ref{gg}) to a system of coupled linear differential equations.   Substituting Eq.~(\ref{lor}) into Eq.~(\ref{g}), we integrate over $E_p$, thus obtaining
\begin{align}
G_{a}(t,t')=g_a^{}(t)g_a^{}(t')\, e^{i(\epsilon_a^{}-i\Lambda_a)(t'-t)}
\label{g11}
\end{align}
where $t'\le t$ and
\begin{align}
g_{a}(t)=\sqrt{{\Gamma_{a}\Lambda_{a}\over2}}\,
w_a(t)\, .
\end{align}
Substituting Eq.~(\ref{g11}) into Eq.(\ref{gg}) and introducing the auxiliary amplitude
\begin{align}
b_{a}^{\prime(\alpha)}(E_{\bbk},t)=
-i\,g_{a}(t)\,\int\limits_0^t e^{i(\epsilon_{a}^{}-E_{\bbk}-i\Lambda_{a})(t'-t)}
b_0^{(a)}(E_{\bbk},t')dt'
\label{gg12}
\end{align}
where $\alpha=L,R$ denotes the reservoir, occupied by electron at $t=0$, we can rewrite Eq.~(\ref{gg}) as a system of coupled equations
\begin{subequations}
\label{g12}
\begin{align}
&{d\over dt}\,b_0^{(\alpha)}(E_{\bbk},t)
=i\big[E_{\bbk}-E_0(t)\big]b_{0}^{(\alpha)}(E_{\bbk},t)
\nonumber\\
&-i\,g_L^{}(t)b_L^{\prime(\alpha)}(E_{\bbk},t)
-i\,g_R^{}(t)b_R^{\prime(\alpha)}(E_{\bbk},t)
-i\Omega^{}_{\bbk}(t)
\label{g12a}\\
&{d\over dt}\,b^{\prime(\alpha)}_{L}(E_{\bbk},t)=
i\big(E_{\bbk}-\epsilon_{L}^{}+i\Lambda_{L}\big)
b^{\prime(\alpha)}_{L}(E_{\bbk},t)
\nonumber\\
&~~~~~~~~~~~~~~~~~~~~~~~~~~~~~~~~~~~~-i\,g_{L}^{}(t)
b^{(\alpha)}_{0}(E_{\bbk},t)
\label{g12b}\\
&{d\over dt}\,b^{\prime(\alpha)}_{R}(E_{\bbk},t)=
i\big(E_{\bbk}-\epsilon_{R}^{}+i\Lambda_{R}\big)
b^{\prime(\alpha)}_{R}(E_{\bbk},t)\nonumber\\
&~~~~~~~~~~~~~~~~~~~~~~~~~~~~~~~~~~~~-i\,g_{R}^{}(t)
b^{(\alpha)}_{0}(E_{\bbk},t)
\label{g12c}
\end{align}
\end{subequations}

Solving these equations, we can evaluate the single electron current, Eq.~(\ref{secur}) and the time-dependent transmissions, Eq.~(\ref{secur}), (\ref{transm1}). For instance, using Eqs.~(\ref{g11}), (\ref{gg12}), we find \begin{align}
T_{L\to R}(E_{\bl}^{},t)=4\pi \varrho_L(E_{\bl})g_R^{}(t)\,{\rm Im}\,\big[q_{0R}^{(L)}(E_{\bl},t)\big]\, ,
\label{secur1}
\end{align}
where
\begin{align}
q_{0R}^{(L)}(E_{\bl},t)
=b_0^{(L)}(E_{\bl},t)b_R^{\prime(L)*}(E_{\bl},t)\, .
\label{offdiag}
\end{align}
The same expression is obtained for $T_{R\to L}(E_{\br},t)$, Eq.~(\ref{transm1b}), with $R\leftrightarrow L$, $\bl\leftrightarrow \br$.

\begin{figure}[h]
\includegraphics[width=8cm]{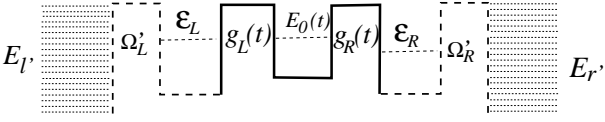}
\caption{(Color online) Quantum dot coupled to two fictitious wells, which incorporate the non-Markovian component of the leads spectrum. These wells are coupled to the effective (Markovian) leads of an infinite band-width.}
\label{fg5}
\end{figure}

Equations~(\ref{g12}) have simple interpretation. They describe electron transport through a triple-dot coupled to two Markovian leads with density of states $\varrho'_{L,R}$, by a coupling $\Omega'_{L,R}$, as shown schematically in Fig.~\ref{fg5}.
Two of the dots (left and right) are fictitious, which account for the non-Markovian (Lorentzian) component of the corresponding lead's spectrum. It implies that the original reservoir basis $|l\rr$, $|r\rr$ is spilt into two components \cite{g1,g2}
\begin{align}
&|l\rr\lr l|=|L\rr\lr L|+\sum_{l'}|l'\rr\lr l'|\, ,
\nonumber\\
&|r\rr\lr r|=|R\rr\lr R|+\sum_{r'}|r'\rr\lr r'|\, .
\label{nbas}
\end{align}

In this basis, the original Hamiltonian, Eq.~(\ref{ham}) can be mapped to the following one, \begin{align}
&H_1(t)=\sum_{l'}E_{l'}^{}\hat c_{l'}^\dagger\hat c_{l'}+\sum_{r'}E_{r'}^{}\hat c_{r'}^\dagger \hat
c_{r'}+\epsilon_L^{}\hat c_L^\dagger \hat c_L^{}+\epsilon_R^{}\hat c_R^\dagger \hat c_R^{}\nonumber\\
&+E_0(t)\hat c_0^\dagger \hat c_0+\Big(g_L^{}(t)\hat c^\dagger_L \hat c_0
+g_R^{}(t)\hat c^\dagger_R \hat c_0
\nonumber\\
&~~~~~~~~~~~~~~
+\sum_{l'}\Omega'_L\hat
c^\dagger_{l'} \hat c_0 +\sum_{r'}\Omega'_R\hat c^\dagger_{r'} \hat c_0+H.c.\Big)\, .
\label{ham1}
\end{align}
One can check that the equation of motion for the wave-function, produced by this Hamiltonian, coincides with Eqs.~(\ref{g12}), obtained from the Hamiltonian (\ref{ham}), with the spectral function (\ref{lor}), providing that
$\pi\Omega_{L,R}^{\prime 2}\varrho'_{L,R}=\Lambda_{L,R}$. The latter can be considered as level-widths of the fictitious dots,  Fig.~\ref{fg5}. Note that the same mapping, as given by Eq.~(\ref{ham1}), has been recently obtained for a rather general case, but using a different method \cite{gernot}.

In the case of time-independent Hamiltonian, $E_0(t)=E_0$ and $w_{L,R}^{}=1$, Eqs.~(\ref{g12}) can be easily solved, in particular in the steady-state limit ($t\to\infty$). Then the l.h.s. of Eqs.~(\ref{g12}) vanishes, so these equations becomes  algebraic. It is useful to introduce the amplitudes $\bar B_0^{(\alpha)}(E_{\bbk})$, defined as
$\Omega_{\alpha}\bar B_{0}^{(\alpha)}(E_{\bbk})=b_0^{(\alpha)}(E_{\bbk},t\to\infty)$ and
$\Omega_{\alpha}\bar B^{\prime(\alpha)}_{L,R}(E_{\bbk})
=b^{\prime(\alpha)}_{L,R}(E_{\bbk},t\to\infty)$. One obtains
\begin{subequations}
\label{g13}
\begin{align}
&
\big(E_{\bbk}-E_0\big)\bar B_{0}^{(\alpha)}
-\,g_L^{}\bar B_L^{\prime(\alpha)}
-\,g_R^{}\bar B_R^{\prime(\alpha)}=1
\label{g13a}\\
&
\big(E_{\bbk}-\epsilon_{L}^{}+i\Lambda_{L}\big)
\bar B^{\prime(\alpha)}_{L}-g_{L}^{}
\bar B^{(\alpha)}_{0}=0
\label{g13b}\\
&
\big(E_{\bbk}-\epsilon_{R}^{}+i\Lambda_{R}\big)
\bar B^{\prime(\alpha)}_{R}-g_{R}^{}
\bar B^{(\alpha)}_{0}=0
\label{g13c}
\end{align}
\end{subequations}
Solving these equations we find
\begin{align}
\bar B_{0}^{(\alpha)}(E_{\bbk})={1\over E_{\bbk}-E_0-{g_L^2\over E_{\bbk}-\epsilon_{L}^{}+i\Lambda_{L}}-{g_R^2\over E_{\bbk}-\epsilon_{R}^{}+i\Lambda_{R}}}
\label{barB0}
\end{align}
where $\bbk\in\alpha$, and
\begin{align}
\bar B_{L,R}^{\prime(\alpha)}(E_{\bbk})={g_{L,R}^{}\over E_{\bbk}-\epsilon_{L,R}^{}+i\Lambda_{L,R}}\bar B_{0}^{(\alpha)}(E_{\bbk})\, .
\label{barB1}
\end{align}

Let us evaluate the steady-state transmission coefficient $\bar T_{L\to R}(E)=T_{L\to R}(E,t\to\infty)$, Eq.~(\ref{secur1}). One finds from Eqs.~(\ref{barB0}), (\ref{barB1}),
\begin{align}
{\rm Im}\,\Big[\bar q_{0R}^{(L)}(E)\Big]
={\Omega_L^2(E)g_R\Lambda_R\over (E-\epsilon_R)^2+\Lambda_R^2}|\bar B_0^{(L)}(E)|^2
\end{align}
where $\bar q_{0R}^{(L)}(E)=q_{0R}^{(L)}(E,t\to\infty)$
Substituting this result into Eg.~(\ref{secur1}) by using Eq.~(\ref{lor}), we finally obtain
\begin{align}
\bar T_{L\to R}(E)={\Lambda_L^2\Gamma_L^{}\Gamma_R^{}\Lambda_R^2|\bar B_0^{(L)}(E)|^2\over \big[E-\epsilon_L^{})^2+\Lambda_L^2\big]
\big[E-\epsilon_R^{})^2+\Lambda_R^2\big]}
\label{fwidth}
\end{align}
The same expression is obtained for $\bar T_{R\to L}(E)$, thus verifying the time-reversal symmetry, Eq.~(\ref{timeind}), for leads of finite band-width. It is easy to find that in the Markovian limit, $\Lambda_{L,R}\to\infty$, Eq.~(\ref{fwidth}) coincides with the Breit-Wigner formula, Eq.~(\ref{brwig}) for $t\to\infty$.

Finally, we point out that the last term in Eq.~(\ref{g12a}) is originated by the initial condition, corresponding to occupied level ($E_{\bbk}$) in one of the leads, Fig.~\ref{fg1}. In the new basis,  Eq.~(\ref{nbas}), it would correspond to a linear superposition of states $|l'\rr$ and $|L\rr$ (or $|r'\rr$ and $|R\rr$). This implies that by solving the time-dependent Scr\"odinger equation for the initial condition, corresponding to occupied state $|l'\rr$ (or $|r'\rr$) in a fictitious reservoir, Fig.~\ref{fg5}, the time-dependence of transmission coefficient $T(E,t)$, would be different. However the steady-state limit of the transmission coefficient, $\bar T(E)$, does not depend on the initial state, and it is always given by Eq.~(\ref{fwidth}).

\section{Time-dependent current of non-interacting electrons}
\label{sec3}

Now we are going to many-electron current, flowing between the leads. Although the electrons are treated as non-interacting particles, it is necessary to consider many-body wave-function to account non-trivial Pauli principle effects, in particular for the time-dependent Hamiltonian \cite{hanggi}. Such a treatment has been done in Refs.~[\onlinecite{single,GAE}], concentrating on the leads at zero temperature. Here we present a general case.

\subsection{Pure and mixed states.}
\label{sec3a}

Consider the leads ($L,R$), Fig.~\ref{fg1}, which are filled at $t=0$ by $N_{L,R}$ electrons, respectively. These numbers vary in time, but the total number of electrons, $N=N_L+N_R+n_0$ remains constant. (Here $n_0=0,1$ denotes number of electrons, initially occupying the quantum dot). In the following we consider the limit of $N_{L,R}\to\infty$.

The wave-function of an entire system, $|\Psi^{(\nu)}_{}(t)\rr$, can be written at $t=0$ as
\begin{align}
|\Psi^{(\nu)}_{}(0)\rr=(c_0^\dagger)^{n^{}_0}
\prod_{\bbk\in\nu}c_{\bbk}^\dagger |0\rangle\ .
\end{align}
where $\nu$ denotes a particular configuration of the reservoirs energy levels ($E_{\bbk}$), occupied by $N_{L,R}$ electrons at $t=0$. For instance, if each of the leads at $t=0$ is taken at zero temperature \cite{GAE}, then $\nu$ comprises the states $\bbk=\{\bar l,\bar r\}$ of the leads, corresponding to $E_{\bar l}\le\mu_L$ and $E_{\bar r}\le\mu_R$, where $\mu_{L,R}$ denote Fermi energies of the leads.

In order to obtain the total many-body wave-function $|\Psi_{}^{(\nu)}(t)\rr$ as a solution of time-dependent Schr\"odinger equation, we use an {\em Ansatz} for $|\Psi_{}^{(\nu)}(t)\rangle$, by taking it as a (Slater) product of single-electron wave functions,
\begin{align}
|\Psi_{}^{(\nu)}(t)\rangle =\prod_{\bbk\in\nu}\hat\Phi_{}^{(\bbk)\dagger}(t)|0\rangle\ ,
\label{a3}
\end{align}
where $\hat\Phi_{}^{(\bbk)\dagger}(t)$ is given by Eq.~(\ref{wf1}), (\ref{c6}) with the initial conditions Eq.~(\ref{incond0}).

If the state of quantum system is described by a wave function, then it is referred to as a ``pure'' state. However, in general, the system can be in a ``mixture'' of pure states, described by the density matrix $\rho(t)$. The latter is obtained from the von Neumann equation
\begin{align}
i\,{d\over dt}\rho(t)=[H(t),\rho (t)]
\label{vn}
\end{align}
uniquely defined by the initial condition
\begin{align}
\rho(0)=\sum_\nu p_\nu|\Psi_{}^{(\nu)}(0)\rr\lr\Psi_{}^{(\nu)}(0)|\, .
\label{incond2}
\end{align}
where $p_\nu$ is a probability of finding the system in a configuration $\nu$. One easily finds that the density matrix
\begin{align}
\rho(t)=\sum_\nu p_\nu |\Psi_{}^{(\nu)}(t)\rr\lr\Psi_{}^{(\nu)}(t)|\, .
\label{mixture}
\end{align}
with $|\Psi_{}^{(\nu)}(t)\rr$ given by Eq.~(\ref{a3}), is indeed the unique solution of Eq.~(\ref{vn}), corresponding to the initial condition (\ref{incond2}).

Thus, instead of solving the matrix equation (\ref{vn}) directly, one can solve the Schr\"odinger equation for a single electron wave-function, Eq.~(\ref{wf1}), for different initial states of the electron ($E_{\bbk}$), corresponding to a set $\nu$. Each of the many-body wave functions of this set is given by the Slater product of the single-electron wave functions, Eq.~(\ref{a3}). Finally, the density matrix of a general mixture state is an incoherent mixture of these pure many-body states ($\nu$), Eq.~(\ref{mixture}).

One can easily realize that $p_\nu$ can be replaced by the {\em initial} electron distribution,  $f_{L/R}^{}(E_{\bbk})$, in the left/right lead, respectively. In the continuous limit
$\int_{-\infty}^\infty f_{L,R}^{}(E)\varrho_{L,R}^{}(E)\,dE=N_{L,R}\to\infty$. For the definiteness  $f_{L,R}^{}(E)$ are represented by Fermi functions, $f(E)=1/[1+e^{(E-\mu)/{\cal T}}]$. However, it can be any other distribution. Note that the steady state implies the same order of limits as in the previous case of a single-electron wave function. In addition the limit $N_{L,R}\to\infty$ takes place before the limit of $t\to\infty$.

\subsection{Charges and currents.}
\label{sec3b}

It was shown in Refs.~[\onlinecite{single,GAE}] that the total charge on the dot at time $t$, $Q_0^{}(t)$, is an {\em incoherent} sum of single-electron probabilities,
Eq.~(\ref{q01}), over all the states $\bbk$, initially occupied by electrons,
\begin{align}
&Q_0^{}(t)=
\sum_{\alpha,\bbk\in\alpha}q_0^{(\alpha)}(E_{\bbk},t)\equiv
Q_0^{(L)}(t)+Q_0^{(R)}(t)+q_0^{(0)}(t)\, .
\label{QLR1}
\end{align}
where $\alpha=L,R,0$. Here we separated the sum over $\bbk=\{\bl,\br,0\}$ into three contributions for electrons coming from the left/right leads and the quantum dot. In continuous limit one can write
\begin{align}
Q_0^{(L,R)}(t)=\int\limits_{-\infty}^\infty q_0^{(L,R)}(E,t)\,f_{L,R}^{}(E)\varrho_{L,R}^{}(E)dE
\label{QLR}
\end{align}
where $E\equiv E_{\bl,\br}$.

Similar to Eqs.~(\ref{QLR1}), (\ref{QLR}), it was shown in Refs.~[\onlinecite{single,GAE}] that the total current in the leads is given by an {\em incoherent} sum of single-electron currents (transmission probabilities), Eqs.~(\ref{secur}), (\ref{transm1}), over all states $\bbk$, initially occupied by electrons,
$I_\alpha^{} (t)=\sum_{\bbk}{\cal I}_\alpha^{} (E_{\bbk},t)$, where $\alpha =L,R$ denotes the lead. For instance, using Eq.~(\ref{opt}), we obtain for the right-lead current
\begin{align}
I_R(t)&=\int\limits_{-\infty}^\infty \big[T_{L\to R}(E,t)f_L(E)
-T_{R\to L}(E,t)f_R(E)\big]{dE\over 2\pi}\nonumber\\
&-\dot Q_0^{(R)}(t)+{\cal I}_R^{(0)}(t)\, ,
\label{irr}
\end{align}
where $E\equiv E_{\bbk}$. The left-lead current, $I_L(t)$, is given by the same formula  with $R\leftrightarrow L$ and by reversing the sign of current. In general, the circuit current is
\begin{align}
I(t)=\beta_R^{} I_R^{}(t)+\beta_L^{} I_L^{}(t)
\label{cc}
\end{align}
where the coefficients $\beta_{L,R}^{}$ with $\beta_L^{}+\beta_R^{}=1$ are depending on a circuit geometry (the junction capacities)  \cite{bb}.

The last term of Eq.~(\ref{irr}), ${\cal I}_R^{(0)}(t)$, is given by Eq.~(\ref{secur}) for $\bbk=0$. It always vanishes for $t\to\infty$ (c.f. Eq.~(\ref{oc2})) and is identically zero if the dot is initially empty, $q_0^{(0)}(0) = 0$.  The second term of Eq.~(\ref{irr}), $\dot Q_0^{(R)}(t)$, represents so-called ``displacement'' current \cite{you}, originated by a retardation of the current flowing  through the dot to the same reservoir.

If the system reaches its steady state limit at $t\to\infty$ the displacement current vanishes. As a result, Eq.~(\ref{irr}) (and Eq.~(\ref{cc})) for the steady-state current, $\overline{I}=I(t\to\infty)$ reads
\begin{align}
\overline{I}=\int\limits_{-\infty}^\infty \big[\bar T_{L\to R}(E)f_L(E)
-\bar T_{R\to L}(E)f_R(E)\big]{dE\over 2\pi}\, .
\label{land}
\end{align}
In the case of time-reversal symmetry, Eq.~(\ref{timeind}), one finds that Eq.~(\ref{land}) becomes the standard Landauer formula. Thus, Eq.~(\ref{irr}) can be considered as a generalization of the Landauer formula to transient currents and time-dependent potentials.

Equation~(\ref{irr}) looks similar to the time-dependent Landauer-type formula, (Eq.~(44) of Ref.~[\onlinecite{hanggi}]), obtained for a periodic drive by using the Floquet approach \cite{hanggi2,hanggi}. In contrast, our results are derived without any use of the Floquet expansion. Hence, despite of its ``scattering'' form, Eq.~(\ref{irr}) is valid for arbitrary time-dependent drive. In addition, our time-dependent transmission probabilities, are directly related to the time-dependent occupation amplitude of a quantum system, given by Eq.~(\ref{gg}). The latter is obtained from a single-particle Schr\"odinger equation with arbitrary initial state at $t=0$ (or at any other time). Therefore we do not need to consider the initial state at $t\to -\infty$ (as in the standard scattering theory). This considerable simplifies the treatment.

In the following, we use the generalized Landauer formula, to study time-dependent current at zero-bias ($f_L(E)=f_R(E)$). Note, that in the case of time-independent Hamiltonian, the time-reversal symmetry of the transmission probability holds. As a result, the steady-state zero-bias current would always vanish, Eq.~(\ref{land}). Below we investigate the zero-bias steady-state current for randomly fluctuating potentials.

\section{Electron current through fluctuating level}
\label{sec4}

Let us return to quantum dot, coupled to two (Markovian) reservoirs, Fig.~\ref{fg1}, described by the Hamiltonian~(\ref{ham}). Now we consider the tunneling couplings as time and energy independent, $\Omega_{L,R}(t)=\Omega_{L,R}$ ($w_{L,R}=1$ in Eq.~(\ref{omtp})), but the energy level of the dot is {\em time-dependent}
\begin{align}
E_0(t)=E_0+\xi(t) {U\over2}\, ,
\label{noise0}
\end{align}
where $\xi(t)=\pm 1$ is jumping randomly from 1 to -1 (or from -1 to 1) at a rate $\gamma_+$ (or $\gamma_-$), independently of its previous history. This represents so-called ``telegraph noise''. We denote $P_\pm^{}(t)$ as probabilities for finding $\xi(t)$ at the values $\xi=\pm 1$ at time $t$, while $P_+^{}(t)+P_-^{}(t)=1$. This quantity is obtained from the corresponding Markovian rate equation
\begin{align}
\dot P_\pm^{}(t)=-\gamma_\pm^{} P_\pm^{}(t)+\gamma_\mp^{} P_\mp^{}(t)=-\gamma P_\pm^{}(t)+\gamma_\mp^{}
\label{rate}
\end{align}
where $\gamma=\gamma_+^{}+\gamma_-^{}$.

Solving Eq.~(\ref{rate}) we find
\begin{align}
P_\pm^{}(t)={\gamma_\mp^{}\over \gamma}+\Big[P_\pm^{}(0)-{\gamma_\mp^{}\over \gamma}\Big]e^{-\gamma t}
\label{rate1}
\end{align}
Thus, in the steady-state limit, the distributions are independent of the initial condition, $\bar P_\pm = P_\pm(t\to\infty)=\gamma_\mp^{}/\gamma$. If the noise is generated by a heat bath of temperature $\cT$ (see for instance, Ref.~[\onlinecite{bergli}]), then
\begin{align}
{\gamma_+^{}\over\gamma_-^{}}=
{\bar P_-^{}\over\bar P_+^{}}=e^{U/\cT}.
\label{relocc}
\end{align}

The average value of $\xi (t)$ is
\begin{align}
\bar\xi =\lr \xi(t)\rr=\sum_{\xi=\pm 1}\bar P_\xi \xi={\gamma_-^{}-\gamma_+^{}\over \gamma}={1-e^{U/\cT}\over 1+e^{U/\cT}}
\label{rate2}
\end{align}
Therefore $\bar\xi=0$ for infinite temperature , ($\gamma_+=\gamma_-=\gamma/2$) and $\bar\xi=-1$ for zero temperature ($\gamma_+=\gamma$, $\gamma_-=0$). The case of infinite temperature has been considered in Ref.~[\onlinecite{GAE}]. Here we consider a finite temperature, $\cT$.

\begin{figure}[h]
\includegraphics[width=8cm]{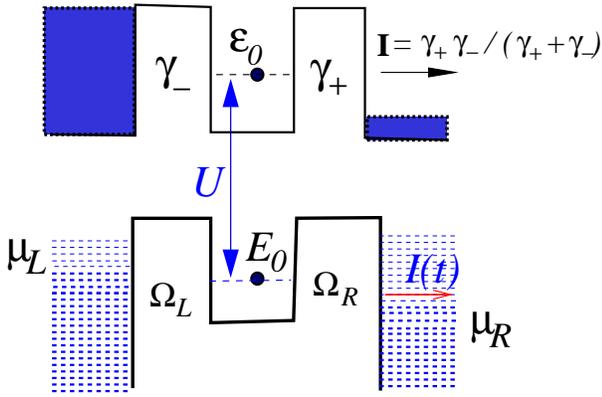}
\caption{(Color online) Two quantum dots are coupled capacitively via Coulomb repulsion $U$. The upper dot is under large bias} \label{fg4}
\end{figure}

In fact, the telegraph noise may not be related to thermal fluctuations. For instance, it can be generated by a fluctuator, representing by a quantum dot at high bias voltage \cite{gm,ensslin,david,sanchez}, (the upper dot in Fig.~\ref{fg4}). There an electron from the left lead enters the dot with tunneling rate $\gamma_-$ and leaves it to the right lead with tunneling rate $\gamma_+$. As a result the charge inside the dot is fluctuating in time, creating fluctuation of the energy level $E_0$ of a nearby quantum dot via the Coulomb interaction $U$.

If the upper dot is under large bias, there is no back action of the lower dot on charge fluctuations inside the upper dot \cite{gm,amnon}. Indeed, if $\epsilon_0$ and $\epsilon_0+U$ are deeply inside the bias,
the level's shift $U$ due to the electron-electron interaction does not affect the charge-correlator of the upper dot. Thus, the effect of the upper dot on the lower one is entirely  accounted for by an external telegraph noise fluctuating the level $E_0$.

An absence of the back action can be verified experimentally by measuring the current {\bf I} in the upper dot, when the lower dot is set at zero bias $\mu_L=\mu_R=\mu$. If the current {\bf I} and its noise spectrum remain the same for $\mu\ll E_0$ (empty dot) and $\mu\gg E_0$ (occupied dot), then the modified Landauer formula,  Eq.~(\ref{irr}), for the quantum-mechanical (ensemble averaged) current, $I(t)$, can be applied.

In the case of noise, Eq.~(\ref{irr}), must include an  additional ensemble average, $T(E,t)\to \lr T(E,t)\rr$ and $Q(t)\to\lr Q(t)\rr$,
over all particular histories of the noise. As follows from Eqs.~(\ref{transm1}), (\ref{QLR}), (\ref{irr}), it can be done by averaging the probability, $q_0^{(\alpha)}(E,t)\to\lr  q_0^{(\alpha)}(E,t)\rr$, given by Eq.~(\ref{oc1}). In fact, it is more convenient to average the corresponding {\em probability amplitude}, $b_0^{(\alpha)}(E,t)$, for finding the dot occupied, Eq.~(\ref{gg1}) with $E_0(t)$ given by Eq.~(\ref{noise0}). It reads
\begin{align}
{d\over dt}b_0^{(\alpha)}(E,t)
=i\left[E-\xi(t) {U\over2}+i{\Gamma\over2}\right]b_0^{(\alpha)}(E,t)
-i\Omega_{\alpha}\, ,
\label{amp}
\end{align}
where $E\equiv E_{\bbk}$ and $\Gamma=2\pi\big(\Omega_L^2\varrho_L+\Omega_R^2\varrho_R\big)$.
(We choose the scale where $E_0=0$.) The average probability
$\lr q_0^{(\alpha)}(E,t)\rr=\lr |b_0^{(\alpha)}(E,t)|^2\rr$, can be determined from Eq.~(\ref{dq0}), which in the wide-band limit, Eq.~(\ref{gal}), becomes
\begin{align}
{d\over dt}\,q_0^{(\alpha)}(E,t) =-\Gamma q_0^{(\alpha)}(E,t)-2\Omega_{\alpha} {\rm Im}[b_0^{(\alpha)}(E,t)]\, ,
\label{q0}
\end{align}

Solving this differential equation and then averaging over the noise we obtain \cite{GAE}
\begin{align}
\langle q_0^{(\alpha)}(E,t)\rangle=-2\Omega^{}_\alpha \int\limits_0^t
e^{\Gamma (t'-t)}{\rm Im}\Big[\langle
b_0^{(\alpha)}(E,t')\rangle\Big] dt'\, .
\label{rate3avsa}
\end{align}
This equation directly relates $\langle q_0^{(\alpha)}(E,t)\rangle$ to $\langle
b_0^{(\alpha)}(E,t)\rangle$. Note that $\langle q_0^{(\alpha)}(E,t)\rangle=\langle |b_0^{(\alpha)}(E,t)|^2\rangle\not =|\lr b_0^{(\alpha)}(E,t)\rr|^2$.

For averaging Eq.~(\ref{amp}) over the noise, we need to evaluate the term $\lr\xi (t)b_0^{(\alpha)}(E,t)\rr$. In order to do it, we  multiply Eq.~(\ref{amp}) by $\xi (t)$, taking into account that $\xi^2(t)=1$. As a result,
\begin{align}
\langle\xi(t){d\over dt}b_0^{(\alpha)}(E,t)\rangle &=i\left[E+i{\Gamma\over2}\right]
\langle\xi (t)b_{0}^{(\alpha)}(E,t)\rangle\nonumber\\
&~-i{U\over2}\langle b_{0}^{(\alpha)}(E,t)\rangle -i\Omega_{\alpha}\bxi\, .
\label{amp2}
\end{align}
where $\bxi =(\gamma_--\gamma_+)/\gamma$, Eq.~(\ref{rate2}).

In the case of exponential noise-correlator, $\lr\xi(t_1)\xi(t_2)\rr\propto\exp (-\gamma |t_1-t_2|)$, one can use the following ``differential formula'', representing an extension of Shapiro and Loginov result \cite{shapiro} (amended for a non-symmetric noise in Appendix~\ref{app1}),
\begin{align}
{d\over dt}\langle \xi(t)R[\xi(t),t]\rangle&=\langle \xi(t){d\over dt}R[\xi(t),t]\rangle -\gamma\,\langle \xi(t) R[\xi(t),t]\rangle
\nonumber\\
&~~~~~~~~~~~~~~~~~+\gamma \bar\xi R[\xi(t),t]
\label{df}
\end{align}
$R[\xi(t),t]$ is an arbitrary functional of the noise. In our case $R[\xi(t),t]\equiv b_0^{(\alpha)}(E,t)$. Substituting Eq.(\ref{df}) into Eq.~(\ref{amp2}) and average over the noise Eq.~(\ref{amp}), we obtain a system of two coupled differential equations for two functions $\langle b_{0}^{(\alpha )}(E,t)\rangle$ and $\langle b_{0\xi}^{(\alpha)}(E,t)\rangle=\langle \xi(t) b_{0}^{(\alpha)}(E,t)\rangle$,
\begin{subequations}
\label{amp1}
\begin{align}
&{d\over dt}\lr{b_0^{(\alpha)}(E,t)\rr}
=i\left(E+i{\Gamma\over2}\right)\lr b_0^{(\alpha)}(E,t)\rr
\nonumber\\
&~~~~~~~~~~~~~~~~~~~~~~~~~~~
-i{U\over2}\lr b_{0\xi}^{(\alpha)}(E,t)\rr -i\Omega_{\alpha}\, ,
\label{amp1a}\\
&{d\over dt}\langle b_{0\xi}^{(\alpha)}(E,t)\rangle =i\Big[E+i\Big({\Gamma\over2} +\gamma\Big)\Big]
\langle b_{0\xi}^{(\alpha)}(E,t)\rangle\nonumber\\
&~~~~~~~~~~~~~~~~~~~
+\Big(\gamma \bar\xi-i{U\over2}\Big)\langle b_{0}^{(\alpha)}(E,t)\rangle-i\Omega_{\alpha} \bar\xi\, .
\label{amp1b}
\end{align}
\end{subequations}

Equations~(\ref{amp1}) represent a non-trivial extension of our previous work \cite{GAE}, considered only symmetric noise, $\bar\xi=0$ (corresponding to infinite noise temperature, $\cT\to\infty$ in Eq.~(\ref{relocc})). These equations can be simplified furthermore by introducing the following variables,
\begin{align}
&b_{0\pm}^{(\alpha)}(E,t)=\lr b_{0}^{(\alpha)}(E,t)\rr\pm \lr b_{0\xi}^{(\alpha)}(E,t)\rr\, ,
\label{amp4}
\end{align}
Then Eqs.~(\ref{amp1}) become
\begin{align}
&{d\over dt}b_{0\pm}^{(\alpha)}(E,t)=
i\left(E\mp{U\over2}+i{\Gamma\over2}\right)b_{0\pm}^{(\alpha)}(E,t)
\nonumber\\
&~~~~~-\gamma_\pm^{}b_{0\pm}^{(\alpha)}(E,t)
+\gamma_\mp^{}b_{0\mp}^{(\alpha)}(E,t) -2i{\gamma_\mp^{}\over \gamma}\,\Omega_{\alpha}\, .
\label{amp5}
\end{align}
Respectively, the probability of finding the dot occupied, averaged over the noise, Eq.~(\ref{rate3avsa}), can be written as
\begin{align}
\langle q_0^{(\alpha)}(E,t)\rangle={1\over2}\big[q_{0+}^{(\alpha)}(E,t)+
q_{0-}^{(\alpha)}(E,t)\big]
\label{amp6}
\end{align}
where $q_{0\pm}^{(\alpha)}(E,t)$ is given by Eq.~(\ref{rate3avsa}) with replacements $q_0^{(\alpha)}(E,t)\to q_{0\pm}^{(\alpha)}(E,t)$ and $b_0^{(\alpha)}(E,t)\to b_{0\pm}^{(\alpha)}(E,t)$. Finally, the time-dependent transmission, averaged over the noise, $\lr T(E,t)\rr$, is given by Eqs.~(\ref{trans3}).

Equations~(\ref{amp5}), (\ref{amp6}) can be considered as describing an electron with a pseudo-spin $1/2$, traveling through a quantum dot, when the electron interacts with a fictitious field. The latter splits the energy level of the dot into two sub-levels, $E_0\to E_0\pm U/2$, depending of direction of the pseudo-spin. The noise produces jumps between these components with rates $\gamma_\pm^{}$, accounted for by the ``loss'' and ``gain'' terms, $-\gamma_\pm^{}\,b_{0\pm}^{(\alpha)}$ and $\gamma_\mp^{}\,b_{0\mp}^{(\alpha)}$, in Eq.~(\ref{amp5}). The last (source) term in this equation accounts the  electron jump from its initial state (in the leads) to the quantum dot.

\subsection{Steady-state limit.}
\label{sec4a}

In steady-state limit, $(d/dt){b_{0\pm}^{(\alpha)}(E,t)}_{t\to\infty}\to 0$, Eq.~(\ref{amp5}) become a system of algebraic equations

\begin{align}
&\left(E\mp{U\over2}+i{\Gamma\over2}\right)\bb_{0\pm}^{(\alpha)}(E)
\nonumber\\
&~~~~~+i\gamma_\pm^{}\bb_{0\pm}^{(\alpha)}(E)
-i\gamma_\mp^{}\bb_{0\mp}^{(\alpha)}(E) =2{\gamma_\mp^{}\over \gamma}\,\Omega_{\alpha}\, .
\label{amp3}
\end{align}
Here we denoted $\ob_{0\pm}^{(\alpha)}(E)\equiv b_{0\pm}^{(\alpha)}(E,t\to\infty)$
Respectively, in the same limit, Eq.~(\ref{q0}), averaged over the noise, becomes
\begin{align}
\oq_0^{(\alpha)}(E)=-{\Omega_{\alpha}\over\Gamma} {\rm Im}[\ob_{0+}^{(\alpha)}(E)]+\ob_{0-}^{(\alpha)}(E)]
\label{qas}
\end{align}
where $\oq_0^{(\alpha)}(E)=\lr q_0^{(\alpha)}(E,t)\rr_{t\to\infty}$.

Finally, using Eqs.~(\ref{trans3}), (\ref{qas}), we find for the steady-state transmission probability, averaged over the noise, $\oT_{L\to R}(E)=\oT_{R\to L}(E)=\oT(E)$
\begin{align}
\oT(E)=-{\Gamma_L\Gamma_R\over\Gamma} {\rm Im}\,\big[\oB_{0+}^{}(E)+\oB_{0-}^{}(E)\big]
\label{transm3}
\end{align}
where $\oB_{0\pm}^{}(E)=\ob_{0\pm}^{(\alpha)}(E)/\Omega_{\alpha}$ is independent of $\alpha$.

Solving Eqs.~(\ref{amp5}), we find
\begin{align}
\oT(E)=-{2\Gamma_L\Gamma_R\over\Gamma}\, {\rm Im}\Big[{E+{U\over2}\bar\xi+i\big({\Gamma\over2}+\gamma\big)\over D_0(E)}\Big]
\label{transm33}
\end{align}
where
\begin{align}
D_0(E)=\Big(E+i{\Gamma+\gamma\over2}\Big)^2+{\gamma^2\over4}
-{U\over2}\Big({U\over2}+i\gamma\bar\xi\Big)
\label{transm34}
\end{align}
and $\bar\xi\equiv\bar\xi(\cT)$, Eq.~(\ref{rate2}), is an average value of the noise.

Substituting Eq.~(\ref{transm33}) into Eq.~(\ref{irr}) and taking into account that the displacement current vanishes in the limit of $t\to\infty$, we arrive to Eq.~(\ref{land}), for the steady-state current. For infinite noise-temperature, $\bar\xi=0$, the transmission coefficient $\bar T$, Eq.~(\ref{transm33}), coincides with that, found in Refs.~[\onlinecite{GAE,YG}]. For zero noise-temperature, $\bar\xi=-1$, Eq.~(\ref{rate2}), one easily finds that Eq.~(\ref{transm33}) reproduces the Breit-Wigner formula, given by Eq.~(\ref{brwig}) for $t\to\infty$ and $E_0=-U/2$. This result is quite expectable, since at zero noise-temperature, the electron always occupies the lower level inside the well.

Thus, we found that the steady-state current through fluctuating energy level is given by the Landauer-type formula, Eq.~(\ref{land}), where the corresponding transmission coefficients, averaged over the noise, are symmetric under the time-reversal, Eq.~(\ref{timeind}). Obviously, the zero bias current would be vanished in this case.

It is interesting that in contrast with the electron current, the energy flux carried by electrons through a single dot, coupled to Markovian leads, is not zero for the asymmetric dot ($\Gamma_L\not=\Gamma_R$). This was obtained in Ref.~[\onlinecite{ora}], by using the Keldysh formalism for time-dependent nonequilibrium Green’s functions \cite{fn5}. Such a difference with the electron current can be understood as a heating of the leads by an external noise via fluctuations of the dot's energy. In this case the energy currents always flow from the dot to the two leads, proportional to the corresponding rates, $\Gamma_{L,R}$ \cite{ora}.

Note, that the absence of zero-bias current through a single-dot in the case of noise is predicted only for the Markovian leads (wide-band limit), adjacent to the drag system,  If the dot is coupled to leads of finite bandwidths, the situation can be different. As we demonstrated above, this corresponds to electron transport through a coupled-dot system, connected to {\em Markovian} leads, Fig.~\ref{fg5}. In this case the time-reversal symmetry of the transmission probability can be broken due to noise. As a result, the zero-bias current would appear. This case will be investigated below.

With respect to experiments on Coulomb-coupled quantum dots \cite{david,sanchez}, Fig.~\ref{fg4}, a non-zero bias current (Coulomb drag) had been observed. We assume that this current is due to non-Markovian leads of the drag subsystem or due to a back-action from the ``drag'' to the ``drive'' subsystems. In the latter case, we anticipate that the drag current would disappear with increase of the bias voltage, applied to the drive system. This can be verified experimentally.

Regarding the theoretical analysis of the Coulomb drag problem by using Markovian Lindblad Master equations  (with time-independent rates) \cite{sanchez,kohler}, the problem is that we need large bias limit ($\Gamma/V\ll 1$) to derive these equations \cite{gur}. Therefore, when the drag system is at zero voltage, the corresponding Lindblad-type Master equations would not be valid, in general. In contrast, the SEA approach, used in our calculation, is valid for any bias, applied for the drag system, whenever the drive system can be replaced by an external noise.

\section{Double-dot system}
\label{sec5}

\begin{figure}[h]
\includegraphics[width=7cm]{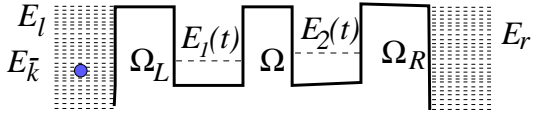}
\caption{(Color online) Resonant tunneling through a double-dot with time-dependent energy levels. A single electron occupies the energy level $E_{\bbk}$ at $t=0$.} \label{fig1}
\end{figure}

Let us consider electron current through a double-dot coupled to two leads, Fig.~\ref{fig1}. The leads are Markovian (infinite band-width). As we demonstrated above, such a system can also correspond to a single dot, coupled to a non-Markovian reservoir (for instance, as shown in Fig.~\ref{fg5} with $\Lambda_L\to\infty$).

Similar to the previous case, Eq.~(\ref{ham}), the system is described by the following tunneling Hamiltonian.
\begin{align}
&H(t)=\sum_lE_l\hat c_l^\dagger\hat c_l+\sum_rE_r\hat c_r^\dagger \hat
c_r+E_1(t)\hat c_1^\dagger \hat c_1+E_2(t)\hat c_2^\dagger \hat c_2
\nonumber\\
&+\left(\Omega \hat c_1^\dagger c_2^{}+\sum_{l}\Omega_L\hat
c^\dagger_l \hat c_0 +\sum_{r}\Omega_R\hat c^\dagger_r \hat c_0+H.c.\right)\, .
\label{a1}
\end{align}
We considered the energy levels inside the dots,  $E_{1,2}(t)$, as time-dependent. However, the tunneling couplings are time-independent.

Let us apply the generalized Landauer formula, Eq.~(\ref{irr}) for the time-dependent current though this system. As in a previous case, we need to evaluate the penetration coefficient, $T(E,t)$. It is done by solving the time-dependent Schr\"odinger equation (\ref{sel}) for a single-electron wave-function, with the initial condition corresponding to occupied energy level $E_{\bbk}$ in one of the leads. Similar to Eq.~(\ref{wf1}), the wave-function can be written as
\begin{align}
|\Phi_{}^{(\bbk)}(t)\rr&=
\Big[\sum_{l}b_{l}^{(\bbk)}(t)c_{l}^\dagger+b_{1}^{(\bbk)}(t)c_1^\dagger
\nonumber\\
&~~~~~~~~~~+b_{2}^{(\bbk)}(t)c_2^\dagger
+\sum_{r}b_{r}^{(\bbk)}(t)\,c_{r}^\dagger\Big]|0\rr\ ,
\label{a4}
\end{align}
where the index $\bbk\equiv\{\bl,\br\}$ denotes the initial condition \cite{fn2},
$b_{l}^{(\bbk)}(0)=\delta^{}_{\bbk\bl}\delta^{}_{{\bar l}l}$ and $b_{r}^{(\bbk)}(0)=\delta^{}_{\bbk\br}\delta^{}_{{\bar r}r}$
(c.f. with Eq.~(\ref{incond0})),

Substituting Eq.~(\ref{a4}) into the Schr\"odinger equation (\ref{sel}). We find the following system of linear equations for the probability amplitudes $b_{}^{(\bbk)}(t)$
\begin{subequations}
\label{a6}
\begin{align}
&i\dot{b}_{l}^{(\bbk)}(t)=E_l\,b_{l}^{(\bbk)}(t)
+\Omega_L\,b_{1}^{(\bbk)}(t)
\label{a6a}\\
&i\dot {b}_{1}^{(\bbk)}(t)=E_1(t)\,b_{1}^{(\bbk)}(t)
+\sum_l\Omega_L\,b_{l}^{(\bbk)}(t)
+\Omega\,b_{2}^{(k)}(t)
\label{a6b}\\
&i\dot {b}_{2}^{(\bbk)}(t)=E_2(t)\,b_{2}^{(\bbk)}(t)
+\sum_r\Omega_R\,b_{r}^{(\bbk)}(t)
+\Omega\,b_{1}^{(\bbk)}(t)
\label{a6c}\\
&i\dot{b}_{r}^{(\bbk)}(t)=E_r\,b_{r}^{(\bbk)}(t)
+\Omega_R\,b_{2}^{(\bbk)}(t)\, .\label{a6d}
\end{align}
\end{subequations}

These equations can be solved in the same way as in the previous case, Eqs.~(\ref{c6}), namely, by resolving
Eqs.~(\ref{a6a}) and (\ref{a6d}) with respect to $b_{l,r}^{(\bbk)}(t)$
\begin{subequations}
\label{a120}
\begin{align}
&b_{l}^{(\bbk)}(t)=e^{-iE_{l}t}\Big[\delta_{\bbk l}\delta_{\bar l l}
-\int\limits_{0}^t
i\,\Omega_{L}\, b_{1}^{(\bbk)}(t')e^{iE_{l}t'}dt'\Big]
\label{a120a}\\
&b_{r}^{(\bbk)}(t)=e^{-iE_{r}t}\Big[\delta_{\bbk r}\delta_{\bar r r}
-\int\limits_{0}^t
i\,\Omega_{R}\, b_{2}^{(\bbk)}(t')e^{iE_{r}t'}dt'\Big]
\label{a120b}
\end{align}
\end{subequations}
and then substituting the result into Eqs.~(\ref{a6b}), (\ref{a6c}). One obtains
\begin{subequations}
\label{aa6c}
\begin{align}
&i\dot {b}_{1}^{(\bbk)}(t)=E_1(t)\,b_{1}^{(\bbk)}(t)+
\Omega_L\delta_{\bbk\bar l}e^{-iE_{\bbk}t}
\nonumber\\
&+\Omega b_2^{(\bbk)}(t')
-i\sum_l\Omega_L^2\int\limits_0^tb_1^{(\bbk)}(t')
e^{iE_l(t'-t)}dt'
\label{a6bb}\\[5pt]
&i\dot {b}_{2}^{(\bbk)}(t)=E_2(t)\,b_{2}^{(\bbk)}(t)+
\Omega_R\delta_{\bbk\bar r}e^{-iE_{\bbk}t}
\nonumber\\
&+\Omega b_1^{(\bbk)}(t')
-i\sum_r\Omega_R^2\int\limits_0^tb_2^{(\bbk)}(t')
e^{iE_r(t'-t)}dt'
\label{a6cc}
\end{align}
\end{subequations}
We emphasise that $\sum_{l,r}$ is extended over {\em all} reservoir states ($E_{l,r}$) without any Pauli principle restrictions.

Now we replace $b_{1,2}^{(\bbk)}(t)=b_{1,2}^{(\alpha)}(E,t) e^{-iEt}$,
(c.f. with Eq.~(\ref{alb})), where $\alpha =\{ L,R\}$ denotes a lead occupied by the electron at $t=0$, and $E= E_k\equiv E_{{\bl},{\br}}$. In continuous limit Eqs.~(\ref{aa6c}) become
\begin{subequations}
\label{a66}
\begin{align}
&\dot b_{1}^{(\alpha)}(E,t)=i\left(E-E_1(t)
+i{\Gamma_L\over2}\right)
\, b_{1}^{(\alpha)}(E,t)\nonumber\\
&~~~~~~~~~~~~~~~~~~~~~~~
-i\,\Omega\, b_{2}^{(\alpha)}(E,t)
-i\Omega_\alpha\,\delta_{\alpha L}\label{a66a}\\
&\dot b_{2}^{(\alpha)}(E,t)=i\left(E-E_2(t)
+i{\Gamma_R\over2}\right)
\, b_{2}^{(\alpha)}(E,t)\nonumber\\
&~~~~~~~~~~~~~~~~~~~~~~~
-i\,\Omega\, b_{1}^{(\alpha)}(E,t)
-i\Omega_\alpha\, \delta_{\alpha R}
\label{a66b}
\end{align}
\end{subequations}
$\Gamma_{L(R)}^{}=2\pi\Omega_{L(R)}^2\varrho_{L(R)}^{}$. Solving these equations we obtain the probability amplitudes for occupation of first and second dot by an electron, coming from left or right reservoir. Then using Eqs.~(\ref{c12p}), (\ref{cali1}) and (\ref{secur}), we obtain for the time-dependent transmission probabilities,  Eqs.~(\ref{transm1}),
\begin{align}
&T_{L\to R}(E,t)=2\pi\varrho_L^{}q_2^{(L)}(E,t)\Gamma_R^{},\nonumber\\
&T_{R\to L}(E,t)=2\pi\varrho_R^{}q_1^{(R)}(E,t)\Gamma_L^{}\, .
\label{transm4}
\end{align}
where $q_{1(2)}^{(\alpha)}(E,t)=|b_{1(2)}^{(\alpha)}(E,t)|^2$ is probability of finding the corresponding dot occupied (c.f. with Eq.~(\ref{trans3})).

Let us take the energy levels of the dots time-independent, $E_{1}(t)=0$ and $E_{2}(t)=\epsilon$. Consider the steady-state limit of Eqs.~(\ref{a66}), corresponding to $(d/dt) b_{1(2)}^{(\alpha)}(E,t\to\infty)\to 0$. Then  Eqs.~(\ref{a6}) becomes a system of algebraic equations that can be easily solved. One finds for the steady-state transmission probabilities
$\bar T(E)=T(E,t\to\infty)$, Eq.~(\ref{transm4})
\begin{align}
\bar T_{L\to R}(E)=\bar T_{R\to L}(E)=\Gamma_L\Gamma_R\left|{\Omega\over D(E)}\right|^2\,
\end{align}
where
\begin{align}
D(E)=\Big(E+i{\Gamma_L\over2}\Big)
\Big(E-\epsilon+i{\Gamma_R\over2}\Big)
-\Omega^2\, .
\end{align}
This result coincides with Eq.~(\ref{fwidth}) when one of the bandwidths, $\Lambda_{L,R}$ is infinity.

\subsection{Telegraph noise.}
\label{sec5a}

Consider now the energy level of the first dot, Fig.~\ref{fig1}, is randomly fluctuating in time, $E_1(t)=(U/2)\xi(t)$, Eq.~(\ref{noise0}), and $E_2(t)=\epsilon$. Then Eqs.~(\ref{a66}), averaged over the noise, read
\begin{subequations}
\label{a9}
\begin{align}
&{d\over dt} \lr b_{1}^{(\alpha)}(E,t)\rr =i\left(E+i{\Gamma_L\over2}\right)
\,\lr b_{1}^{(\alpha)}(E,t)\rr\nonumber\\
&-i{U\over2}\lr \xi(t) b_{1}^{(\alpha)}(E,t)\rr
-i\,\Omega\, \lr b_{2}^{(\alpha)}(E,t)\rr
-i\,\Omega_\alpha\, \delta_{\alpha L}\label{a9a}\\
&{d\over dt}\lr b_{2}^{(\alpha)}(E,t)\rr =i\left(E-\epsilon+i{\Gamma_R\over2}\right)
\, \lr b_{2}^{(\alpha)}(E,t)\rr\nonumber\\
&~~~~~~~~~~~~~~~~~~~~~~~
-i\,\Omega\, \lr b_{1}^{(\alpha)}(E,t)\rr
-i\,\Omega_\alpha\,\delta_{\alpha R}\, .
\label{a9b}
\end{align}
\end{subequations}

In order to evaluate the average $\lr \xi(t) b_{1}^{(\alpha)}(E,t)\rr$ we multiply Eqs.~(\ref{a66}) by $\xi(t)$, taking into account that $\xi^2(t)=1$. As in the case of single dot, we apply the Shapiro-Loginov differential formula Eq.~(\ref{df}), for the terms $\lr \xi(t){d\over dt} b_{1(2)}^{(\alpha)}(E,t)\rr$. Finally we obtain (c.f. with Eqs.~(\ref{amp1})),
\begin{subequations}
\label{a10}
\begin{align}
&{d\over dt} \lr b_{1\xi}^{(\alpha)}(E,t)\rr =i\left(E+i{\Gamma_L+2\gamma\over2}\right)
\,\lr b_{1\xi}^{(\alpha)}(E,t)\rr\nonumber\\
&
+\Big(\gamma \bar\xi -i{U\over2}\Big)\lr b_{1}^{(\alpha)}(E,t)\rr
-i\,\Omega\, \lr b_{2\xi}^{(\alpha)}(E,t)\rr
-i\Omega_\alpha\, \bar\xi\,\delta_{\alpha L}
\label{a10a}\\
&{d\over dt}\lr b_{2\xi}^{(\alpha)}(E,t)\rr =i\left(E-\epsilon+i{\Gamma_R+2\gamma\over2}\right)
\, \lr b_{2\xi}^{(\alpha)}(E,t)\rr\nonumber\\
&
-i\,\Omega\, \lr b_{1\xi}^{(\alpha)}(E,t)\rr+\gamma \bar\xi \langle b_{2}^{(\alpha)}(E,t)\rangle-i\Omega_\alpha\, \bar\xi\,\delta_{\alpha R}
\label{a10b}
\end{align}
\end{subequations}
where $b_{1\xi(2\xi)}^{(\alpha)}(E,t)=\xi(t) b_{1(2)}^{(\alpha)}(E,t)$.

\begin{figure}[tbh]
\includegraphics[width=7cm]{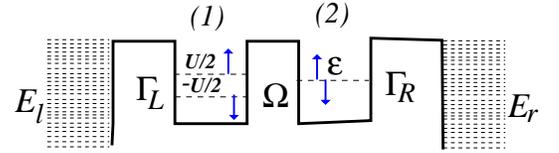}
\caption{(Color online) Resonant tunneling through a double-dot, where the energy level of the first (left) dot is split into two sub-levels, $\pm U/2$, corresponding to the pseudo-spin up and down.} \label{fig2}
\end{figure}

Similar to the previous case, Eq.~(\ref{amp5})), it is useful to introduce the new variables,
\begin{align}
&b_{1\pm}^{(\alpha)}(E,t)=\lr b_{1}^{(\alpha)}(E,t)\rr\pm \lr b_{1\xi}^{(\alpha)}(E,t)\rr\nonumber\\
&b_{2\pm}^{(\alpha)}(E,t)=\lr b_{2}^{(\alpha)}(E,t)\rr\pm \lr b_{2\xi}^{(\alpha)}(E,t)\rr
\label{a13a}
\end{align}
that transform Eqs.~(\ref{a9}), (\ref{a10}) to more transparent equations, which are simpler for treatment. These equations read
\begin{subequations}
\label{a14}
\begin{align}
&{d\over dt}b_{1\pm}^{(\alpha)}(E,t)=i\left(E\mp{U\over2}+i{\Gamma_L\over2}\right)
\,b_{1\pm}^{(\alpha)}-\i\,\Omega b_{2\pm}^{(\alpha)}
\nonumber\\
&~~~~~~~~~~~~~~
-\gamma_\pm^{}\,b_{1\pm}^{(\alpha)}
+\gamma_\mp^{}\,b_{1\mp}^{(\alpha)}
-2i\,\Omega_\alpha{\gamma_\mp^{}\over\gamma}\,
\delta_{\alpha L}
\label{a14a}\\
&{d\over dt}b_{2\pm}^{(\alpha)}(E,t)=i\left(E-\epsilon +i{\Gamma_R\over2}\right)
\,\bb_{2\pm}^{(\alpha)}-\Omega b_{1\pm}^{(\alpha)}
\nonumber\\
&~~~~~~~~~~~~~~
-\gamma_\pm^{}\,b_{2\pm}^{(\alpha)}
+\gamma_\mp^{}\,b_{2\mp}^{(\alpha)}
-2i\,\Omega_\alpha{\gamma_\mp^{}\over\gamma}\,
\delta_{\alpha R}
\label{a14b}
\end{align}
\end{subequations}
As in the previous case of a single dot, Eqs.~(\ref{a14}) can be considered as describing electron with a pseudo-spin $1/2$, travelling through a coupled-dot system and interacting with a fictitious field inside the first (left) dot. This field  splits the first dot level into the two sub-levels, $\pm U/2$, Fig.~\ref{fig2}, where the noise produces jumps between these sub-levels  with rates $\gamma_\pm^{}$.

Solving Eqs.~(\ref{a14}) we obtain the probability amplitudes of occupation the dots, averaged over the noise
\begin{align}
\lr b_{1(2)}^{(\alpha)}(E,t)\rr={1\over2}
\big[b_{1+(2+)}^{(\alpha)}(E,t)+ b_{1-(2-)}^{(\alpha)}(E,t)\big]
\label{a14p}
\end{align}
Next, we have to average the corresponding transmission probabilities, Eq.~(\ref{transm4}), given by a bi-linear form (density-matrix) of these amplitudes, $\lr q_{1(2)}^{(\alpha)}(E,t)\rr =\lr |b_{1(2)}^{(\alpha)}(E,t)|^2\rr$. Since $\lr |b_{1(2)}^{(\alpha)}(E,t)|^2\rr\not =\lr q_{1(2)}^{(\alpha)}(E,t)\rr^2$, we cannot use directly Eqs.~(\ref{a14}), (\ref{a14p}) to evaluate this quantity. For this reason, we derive differential equations, relating the density-matrix of the double-dot system, averaged over the noise, to the averaged amplitudes $\lr b_{1(2)}^{(\alpha)}(E,t)\rr$, similar to Eq.~(\ref{q0}). We find the following equations for the noise-averaged density matrix,
\begin{align}
&\lr q_{1(2)}^{(\alpha)}(E,t)\rr=\lr |b_{1(2)}^{(\alpha)}|^2\rr={1\over2}
\big[q_{1+(2+)}^{(\alpha)}+ q_{1-(2-)}^{(\alpha)}\big]\nonumber\\
&\lr q_{12}^{(\alpha)}(E,t)\rr=\lr b_1^{(\alpha)}b_2^{(\alpha)*}\rr=
{1\over2}
\big[q_{12+}^{(\alpha)}+ q_{12-}^{(\alpha)}\big]
\label{qaver}
\end{align}
(c.f. with Eq.~(\ref{amp6})), where
\begin{subequations}
\label{a15}
\begin{align}
&{d\over dt}q_{1\pm}^{(\alpha)}(E,t)=-\Gamma_L\, q_{1\pm}^{(\alpha)}+i\Omega \big[q_{12\pm}^{(\alpha)}-q_{21\pm}^{(\alpha)}\big]
\nonumber\\
&~~~~~~~~~
-\gamma_\pm^{}q_{1\pm}^{(\alpha)}
+\gamma_\mp^{}q_{1\mp}^{(\alpha)}
-2\Omega_\alpha {\rm Im}[b_{1\pm}^{(\alpha)}]\,\delta_{\alpha L}\label{a15a}\\
&{d\over dt}q_{2\pm}^{(\alpha)}(E,t)=-\Gamma_R\, q_{2\pm}^{(\alpha)}-i\Omega \big[q_{12\pm}^{(\alpha)}-q_{21\pm}^{(\alpha)}\big]
\nonumber\\
&~~~~~~~~~
-\gamma_\pm^{}q_{2\pm}^{(\alpha)}
+\gamma_\mp^{}q_{2\mp}^{(\alpha)}
-2\Omega_\alpha {\rm Im}[b_{2\pm}^{(\alpha)}]\,\delta_{\alpha R}
\label{a15b}\\
&{d\over dt}q_{12\pm}^{(\alpha)}(E,t)=i\Big(\epsilon\mp{U\over2}
+i{\Gamma\over2}\Big)q_{12\pm}^{(\alpha)}
+i\Omega(q_{1\pm}^{(\alpha)}-q_{2\pm}^{(\alpha)})
\nonumber\\
&
-\gamma_\pm^{}q_{12\pm}^{(\alpha)}
+\gamma_\mp^{}q_{12\mp}^{(\alpha)}
-i\Omega_{\alpha}
\big(b_{2\pm}^{(\alpha)*}\delta_{\alpha L}-b_{1\pm}^{(\alpha)}\delta_{\alpha R}\big)\label{a15c}
\end{align}
\end{subequations}
One finds that for each direction of the pseudo-spin, Eqs.~(\ref{a15}) represent the Lindblad-type Master equation for a single electron in a double-dot level system, coupled to leads. The noise only generates transitions between the two pseudo-spin directions. Detailed derivation of Eqs.~(\ref{a15}) is presented in Appendix \ref{app2}.

Solving Eqs.~(\ref{a14}), (\ref{a15}), we find the time-dependent average current, Eqs.~(\ref{transm1}), (\ref{irr}), through a douple-dot system. Now we  concentrate on steady-state.

\subsection{Steady-state.}
\label{sec5b}

Consider Eqs.~(\ref{a14})-(\ref{a15}) in steady-state limit, corresponding to $t\to\infty$. Since in this limit the l.h.s of these equations vanishes, these turn into the following linear algebraic equations
\begin{subequations}
\label{a16}
\begin{align}
&\left(E\mp{U\over2}+i{\Gamma_L\over2}\right)
\,\oB_{1\pm}^{(\alpha)}-\Omega \oB_{2\pm}^{(\alpha)}
\nonumber\\
&~~~~~~~~~~~~~~
+i\gamma_\pm^{}\oB_{1\pm}^{(\alpha)}
-i\gamma_\mp^{}\oB_{1\mp}^{(\alpha)}
=2{\gamma_\mp^{}\over\gamma}
\delta_{\alpha L}
\label{a16a}\\[5pt]
&\left(E-\epsilon +i{\Gamma_R\over2}\right)
\,\oB_{2\pm}^{(\alpha)}-\Omega \oB_{1\pm}^{(\alpha)}
\nonumber\\
&~~~~~~~~~~~~~~
+i\gamma_\pm^{}\oB_{2\pm}^{(\alpha)}
-i\gamma_\mp^{}\oB_{2\mp}^{(\alpha)}
=2{\gamma_\mp^{}\over\gamma}
\delta_{\alpha R}
\label{a16b}
\end{align}
\end{subequations}
and
\begin{subequations}
\label{a17}
\begin{align}
&\Gamma_L\, \oQ_{1\pm}^{(\alpha)}-i\Omega \big[\oQ_{12\pm}^{(\alpha)}-\oQ_{21\pm}^{(\alpha)}\big]
+\gamma_\pm^{}\oQ_{1\pm}^{(\alpha)}
-\gamma_\mp^{}\oQ_{1\mp}^{(\alpha)}\nonumber\\
&~~~~~~~~~~~~~~~~~~~~~~~~~~~~~~~~~
=-2{\rm Im}[\oB_{1\pm}^{(\alpha)}]\delta_{\alpha L}\label{a17a}\\
&\Gamma_R\, \oQ_{2\pm}^{(\alpha)}+i\Omega
\big[\oQ_{12\pm}^{(\alpha)}-\oQ_{21\pm}^{(\alpha)}\big]
+\gamma_\pm^{}\oQ_{2\pm}^{(\alpha)}
-\gamma_\mp^{}\oQ_{2\mp}^{(\alpha)}\nonumber\\
&~~~~~~~~~~~~~~~~~~~~~~~~~~~~~~~~~
=-2{\rm Im}[\oB_{2\pm}^{(\alpha)}]\delta_{\alpha R}
\label{a17b}\\
&\Big(\epsilon\mp{U\over2}
+i{\Gamma\over2}\Big)\oQ_{12+}^{(\alpha)}
+\Omega(\oQ_{1\pm}^{(\alpha)}-\oQ_{2+}^{(\alpha)})
\nonumber\\
&
+i\gamma_\pm^{}\oQ_{12\pm}^{(\alpha)}
-i\gamma_\mp^{}\oQ_{12\mp}^{(\alpha)}
=\oB_{2+}^{(\alpha)*}\delta_{\alpha L}-\oB_{1+}^{(\alpha)}\delta_{\alpha R}\label{a17c}
\end{align}
\end{subequations}
where $\Omega_{\alpha}\oB_{1\pm(2\pm)}^{(\alpha)}(E)
=b_{1\pm(2\pm)}^{(\alpha)}(E,t\to\infty )$ and
$\Omega_{\alpha}^2\oQ_{1\pm(2\pm)}^{(\alpha)}(E)
=q_{1\pm(2\pm)}^{(\alpha)}(E,t\to\infty )$. Solving Eqs.~(\ref{a16}), (\ref{a17}), we obtain the steady-state transmission coefficients, $\oT (E)=T(E,t\to\infty)$, Eq.~(\ref{transm4}), given by the following expressions
\begin{align}
&\oT_{L\to R}(E)=\Gamma_L^{}\,\oQ_{2}^{(L)}
\,\Gamma_R=-i\Omega\big[\oQ_{12}^{(L)}
-\oQ_{21}^{(L)}\big]\Gamma_L^{}
\nonumber\\
&\oT_{R\to L}(E)=\Gamma_R^{}\,\oQ_{1}^{(R)}
\,\Gamma_L=i\Omega\big[\oQ_{12}^{(R)}
-\oQ_{21}^{(R)}\big]\Gamma_R^{}
\label{Tnd}
\end{align}
where $\oQ_{1(2)}^{(\alpha)}(E)
={1\over2}\big[\oQ_{1+(2+)}^{(\alpha)}(E)
+\oQ_{1-(2-)}^{(\alpha)}(E)\big]$.

Equations~(\ref{Tnd}) show that the transmission probability to a reservoir is given by occupation of a quantum dot, adjacent to that reservoir and multiplied by the corresponding tunneling rate (c.f. with Eqs.~(\ref{trans3})). However for a double-dot it can be rewritten through the imaginary part of the off-diagonal occupation density, $\oQ_{12}^{}$, Eqs.~(\ref{Tnd}). This relation is very remarkable. It shows that the transmission probability is given by the off-diagonal density-matrix of an electron in two dots.

Indeed, as follows from Schr\"odinger equation, the quantum transport between two isolated levels always proceeds through linear superposition of these states (Rabi oscillations). This is very different from the transport between two reservoirs with continuum states, or from an isolated to continuum states. Then the tracing over continuum states in the equation of motions, eliminates coherent transitions, making all the transport incoherent, see Ref.~[\onlinecite{gur}] and additional discussion in Sec.~\ref{sec6}.

Using Eqs.~(\ref{Tnd}) we obtain for the asymmetry between the left-right and right-left transmission probabilities, $\Delta \oT(E)= \oT_{L\to R}(E)-\oT_{R\to L}(E)$, the following expression,
\begin{align}
\Delta T(E)=-2\Omega\,{\rm Im}\Big[\Gamma_L\,\oQ_{12}^{(L)}(E)+\Gamma_R\,\oQ_{12}^{(R)}(E)\Big]\, ,
\label{deltaT}
\end{align}
where $\oQ_{12}^{(\alpha)}(E)
={1\over2}\big[\oQ_{12+}^{(\alpha)}(E)
+\oQ_{12-}^{(\alpha)}(E)\big]$.
If $\Delta T(E)\not =0$, then the time-reversal symmetry of  transmission coefficients is broken, leading to the zero-bias current. First we investigate this point  analytically.

\subsection{Analytical treatment.}
\label{sec5c}

Let us take for simplicity symmetric double-dot, $\Gamma_L=\Gamma_R=\Gamma/2$, $\epsilon =0$ and also  symmetric noise $\gamma_+=\gamma_-=\gamma/2$, corresponding to infinite noise temperature, Eq.~(\ref{relocc}).  Although Eqs.~(\ref{a16}), (\ref{a17}) are linear algebraic equations, their  analytical solution looks rather complicated. Therefore in order to obtain simple analytical expressions, we  expand $\oB_{}^{(\alpha)}$ and $\oQ_{}^{(\alpha)}$ in  powers of $\gamma$ by keeping only linear terms. Thus, we represent these variables as
\begin{subequations}
\label{expan1}
\begin{align}
&\oB_{1\pm(2\pm)}^{(\alpha)}=\tB_{1\pm(2\pm)}^{(\alpha)} +\gamma X_{1\pm(2\pm)}^{(\alpha)}\label{expan1a}\\
&\oQ_{1\pm(2\pm)}^{(\alpha)}=\tQ_{1\pm(2\pm)}^{(\alpha)}+\gamma Y_{1\pm(2\pm)}^{(\alpha)}\label{expan1b}\\
&\oQ_{12\pm}^{(\alpha)}=\tQ_{12\pm}^{(\alpha)}+\gamma Y_{12\pm}^{(\alpha)}
\label{expan1c}
\end{align}
\end{subequations}
where $\tB_{1\pm(2\pm)}^{(\alpha)}$, $\tQ_{1\pm(2\pm)}^{(\alpha)}$ and $\tQ_{12\pm}^{(\alpha)}$ are obtained from Eqs.~(\ref{a16}), (\ref{a17}) for $\gamma=0$. One easily finds
\begin{align}
&\tB_{2\pm}^{(L)}=\tB_{1\pm}^{(R)}={\Omega\over D_\pm(E)},~~
\tB_{1\pm}^{(L)}={E+i{\Gamma\over4}\over D_{\pm}(E)}\nonumber\\
&\tB_{2\pm}^{(R)}={E\pm{U\over2}+i{\Gamma\over4}\over D_{\pm}(E)}
\label{BB1}
\end{align}
where
\begin{align}
D_\pm(E)=\Big(E\pm{U\over2}+i{\Gamma\over4}\Big)
\Big(E
+i{\Gamma\over4}\Big)-\Omega^2
\label{BB2}
\end{align}
Note, that the limit of $\gamma=0$ describes the electron  motion through the upper or lower level ($\pm U/2$) of the first dot without the noise, Fig.~\ref{fig2}. Therefore in this case
\begin{align}
\tQ_{1\pm(2\pm)}^{(\alpha)}=|\widetilde B_{1\pm(2\pm)}^{(\alpha)}|^2, ~~\tQ_{12\pm}^{(\alpha)}= \widetilde B_{1\pm}^{(\alpha)}\widetilde B_{2\pm}^{(\alpha)*}
\label{BB3}
\end{align}
The terms $X_{1\pm(2\pm)}^{(\alpha)}$, Eq.~(\ref{expan1a}), are obtained straightforwardly from Eqs.~(\ref{a16}) by taking the limit
\begin{align}
\left. X_{1\pm(2\pm)}^{(\alpha)}=\big[(\oB_{1\pm(2\pm)}^{(\alpha)}
-\tB_{1\pm(2\pm)}^{(\alpha)})/\gamma\big]\right|_{\gamma\to 0}
\end{align}

Substituting Eq.~(\ref{expan1c}) into Eqs.~(\ref{deltaT}) and taking into account that $\Delta T(E)=0$ for $\gamma=0$, we find
\begin{align}
\Delta T(E)=\gamma\,\Omega\,\Gamma\,{\rm Im}\, \big[Y_{12}^{(L)}+Y_{12}^{(R)}\big]
\label{deltaT2}
\end{align}
where $Y_{12}^{(\alpha)}=[Y_{12+}^{(\alpha)}+Y_{12-}^{(\alpha)}]/2$ and $Y_{12\pm}^{(\alpha)}$ are determined from the  equations
\begin{subequations}
\label{a69}
\begin{align}
&\Gamma\, Y_{1\pm}^{(\alpha)}+4\Omega\,{\rm Im}[Y_{12\pm}^{(\alpha)}]=\mp\Delta\tQ_{1}^{(\alpha)}
-4\,{\rm Im}[X_{1\pm}^{(\alpha)}]\delta_{\alpha L}
\label{a69a}\\
&\Gamma\, Y_{2\pm}^{(\alpha)}-4\Omega\,{\rm Im}[Y_{12\pm}^{(\alpha)}]=\mp\Delta\tQ_{2}^{(\alpha)}
-4\,{\rm Im}[X_{2\pm}^{(\alpha)}]\delta_{\alpha R}
\label{a69b}\\
&\Big(\mp{U\over2}+i{\Gamma\over2}\Big)Y_{12\pm}^{(\alpha)}
+\Omega(Y_{1\pm}^{(\alpha)}-Y_{2\pm}^{(\alpha)})\nonumber\\
&~~~~~~~~~~~~~~
=\mp{i\over2}\Delta \tQ_{12}^{(\alpha)}
+X_{2\pm}^{(\alpha)*}\delta_{\alpha L}+X_{1\pm}^{(\alpha)}\delta_{\alpha R}
\label{a69c}
\end{align}
\end{subequations}
with $\Delta\tQ_{1(2)}^{(\alpha)}
=\tQ_{1+(2+)}^{(\alpha)}-\tQ_{1-(2-)}^{(\alpha)}$ and  $\Delta\tQ_{12}^{(\alpha)}
=\tQ_{12+}^{(\alpha)}-\tQ_{12-}^{(\alpha)}$. Solving Eqs.~(\ref{a69}) we arrive to a simple analytical formula for $\Delta T(E)$, Eq.~(\ref{deltaT2}). It becomes even more simple when we expand it in powers of $U$, keeping the leading term $\propto U^2$. We find
\begin{align}
\Delta T(E)={2\gamma\,\Gamma U^2\Omega^2\over\Gamma^2+16\Omega^2}&\,
{E^4+E^2{\Gamma^2\over4}-{1\over16}
\big({\Gamma^2\over4}+4\Omega^2\big)^2\over
\big[\big(E^2+{\Gamma^2\over16}-\Omega^2\big)^2
+{\Gamma^2\over4}\Omega^2\big]^2}\nonumber\\
&~~~~~~~~~~~~~~~~~~
+ {\cal O}\,[\gamma^2, U^4]
\label{dT11}
\end{align}
Let us use Eq.~(\ref{dT11}) for an analysis of the zero bias current.

First, this result confirms that time-reversal symmetry of the transmission coefficient is broken when the energy level of one of the dots is randomly fluctuating, even though, the system by itself is symmetric. In addition, one finds from Eq.~(\ref{dT11}), that $\Delta T(E)$ changes its sign with increase of the energy. This happens at  $E= \pm\Omega$ (for $\Gamma\ll\Omega$), which are the energy eigenvalues of a symmetric double-dot.

The change of sign $\Delta T(E)$ results in changing direction of the zero-bias current, $I_{zb}$ given by the generalized Landauer formula, Eq.~(\ref{land}),
\begin{align}
I_{zb}=\int\limits_{-\infty}^\infty
\Delta T(E)f(E){dE\over 2\pi}\, ,
\label{land1}
\end{align}
where $f(E)$ is Fermi-function of reservoirs. We can evaluate it analytically when the leads are (initially)  at zero temperature, $f(E)=\theta (\mu-E)$, where $\mu$ in the Fermi energy. Using  Eq.~(\ref{dT11}) one finds
\begin{align}
I_{zb}&={\gamma\Gamma U^2\over 16 \pi\big({\Gamma^2\over4}+4\Omega^2\big)}
\left[{\mu\big(\mu^2+{\Gamma^2\over16}-3\Omega^2\big)\over \big(\mu^2+{\Gamma^2\over 16^2}-\Omega^2\big)^2
+{\Gamma^2\over4}\Omega^2}\right.\nonumber\\
&\left.+{1\over\Omega}{\rm Im}\Big[\arctan\Big({4\mu\over \Gamma+4i\Omega}\Big)\Big]\right]+ {\cal O}\,[\gamma^2, U^4]
\label{land2}
\end{align}

It follows from this equation that the zero bias current changes its sign at $\mu=0$. This implies that in the case of Coulomb drag (Fig.~\ref{fg4}), the drag and drive currents would be of opposite directions. Then with increase (decrease) of $\mu$, the current reaches its maximal value, when the Fermi energy is close to the energy level of the isolated double-dot, $\pm\Omega$. Finally, $I_{zb}$ vanishes when $\mu\to\infty$,  Eq.~(\ref{land2}). The same takes place when the leads are at infinite temperature, as follows from  Eq.~(\ref{land1}).

\subsection{Numerical results.}
\label{sec5d}

\begin{figure}[h]
\includegraphics[width=8.5cm]{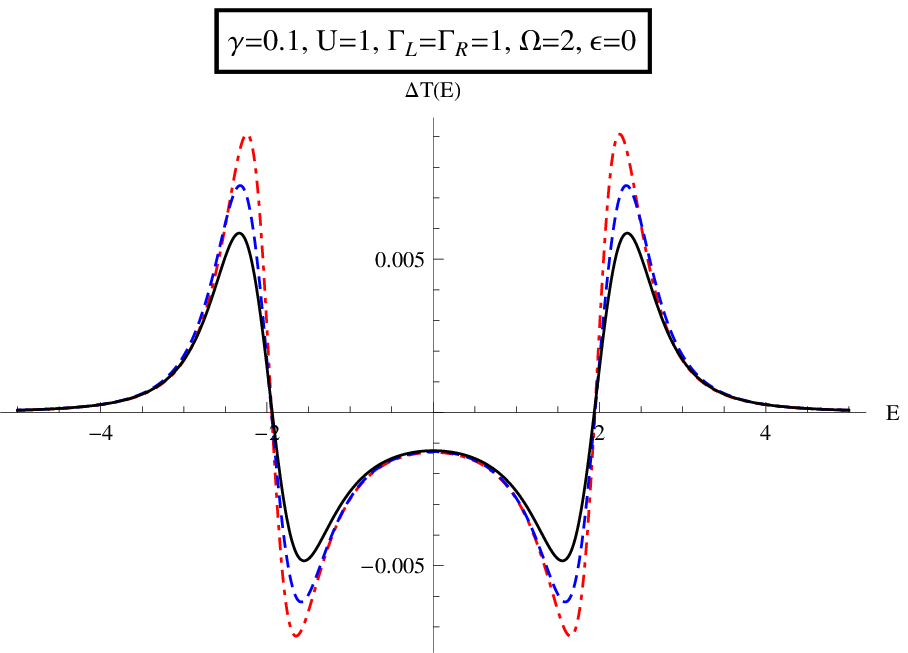}\\
\includegraphics[width=8.5cm]{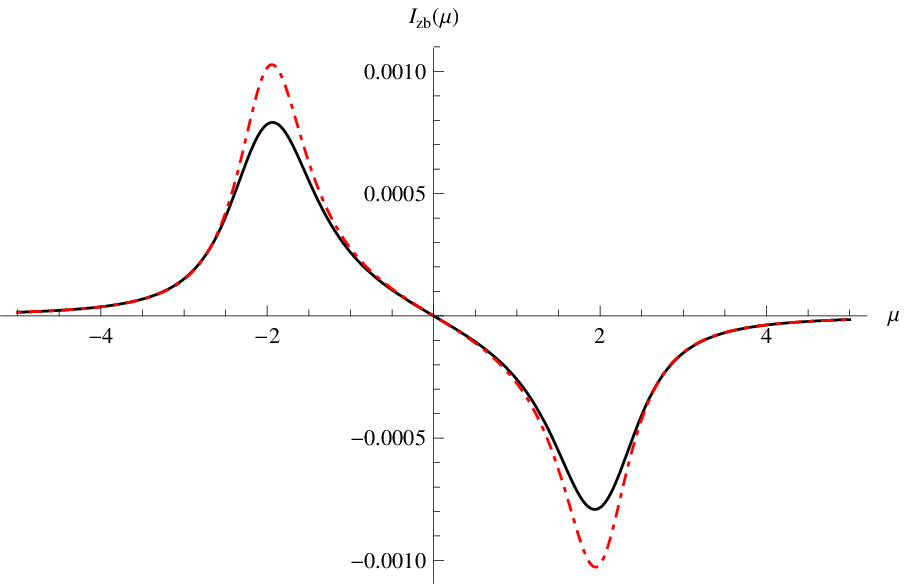}
\vspace{10pt}
\caption{(Color online) Break the time-reversal symmetry of transmission coefficient, $\Delta T$, (upper panel) and zero-bias current, $I_{zb}$, (lower panel) for a symmetric double-dot. Solid line (black) shows exact calculations, where dashed lined (blue) corresponds to leading terms of expansions in powers of $\gamma$. The dot-dashed lines (red) display Eqs.~(\ref{dT11}), (\ref{land2}). }
\label{fig4}
\end{figure}

Let us compare our analytical formulaes Eqs.~(\ref{dT11}), (\ref{land2}) with exact numerical calculations of $\Delta T(E)$, Eq.~(\ref{deltaT}) and  $I_{zb}$, Eq.~(\ref{land1}). The results are shown in Fig.~\ref{fig4} for symmetric double-dot, $\Gamma_L=\Gamma_R=\Gamma/2$, $\epsilon=0$ and symmetric noise, $\cT=\infty$ in Eq.~(\ref{relocc}), corresponding to  $\gamma_+=\gamma_-=\gamma/2$. The upper panel displays $\Delta T(E)$ as a function of electron energy (in arbitrary units), and the lower panel displays the zero-bias current, $I_{zb}(\mu )$, as a function of the Fermi energy ($\mu$). The leads are at zero temperature.
The solid line (black), corresponds to exact numerical solution of Eqs.~(\ref{a16}), (\ref{a17}), where  dashed line (blue) shows approximate calculations (for small $\gamma$), obtained from  Eqs.~(\ref{a69}). The dot-dashed line (red) corresponds to Eqs.~(\ref{dT11}) and (\ref{land2}).

\begin{figure}[h]
\includegraphics[width=8.5cm]{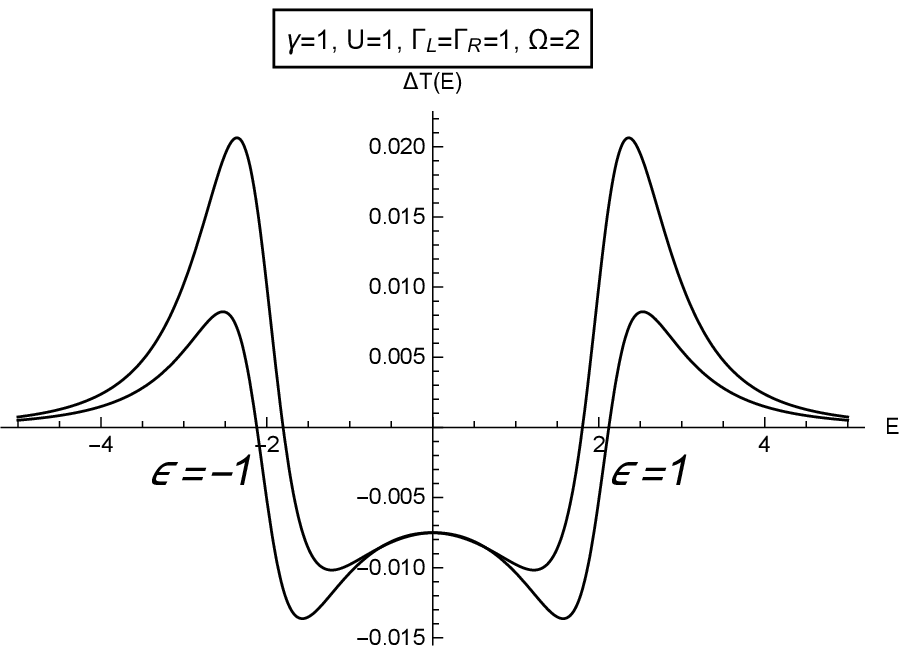}\\
\vspace{10pt}
\includegraphics[width=8.5cm]{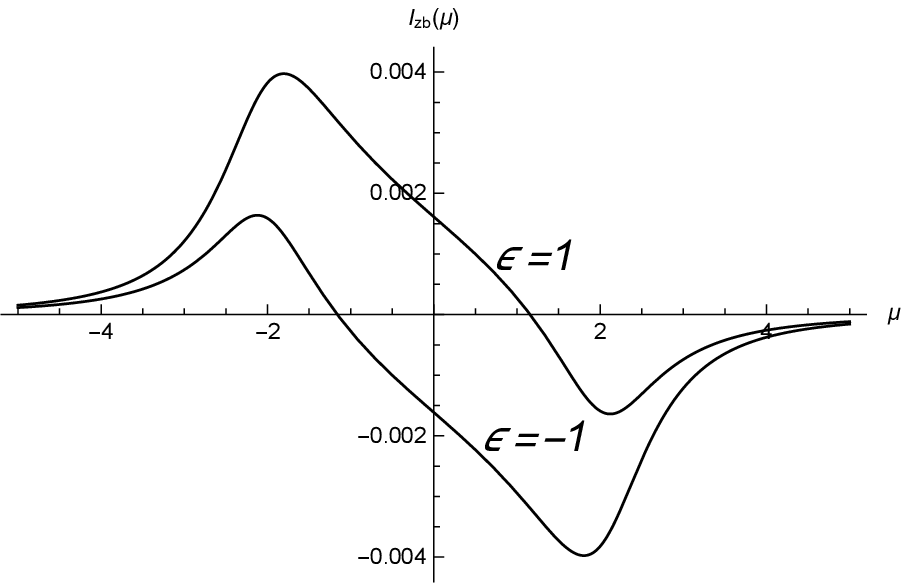}
\vspace{10pt}
\caption{(Color online) Break the time-reversal symmetry of transmission coefficient and zero-bias current in the case of asymmetric double-dot. The upper and lower panels show  $\Delta T(E)$ and $I_{zb}$, respectively,  for $\epsilon=\pm 1$ and $\gamma=1$ (in arbitrary units. Other parameters are the same as in Fig.~\ref{fig4}.}
\label{fig5}
\end{figure}

One finds from Fig.~\ref{fig4} that simple formulaes (\ref{dT11}), (\ref{land2}) reproduce general behavior of $\Delta T$ and zero-bias current very well, in  particular, change of the current direction and  position of its extremal values. A similar behavior of $\Delta T(E)$ and $I_{zb}(\mu )$ is found for a non-symmetric double-dot ($\epsilon\not =0$), and even if the noise spectral width ($\gamma$) is not narrow. It is displayed in  Fig.~\ref{fig5}, which shows the asymmetry of transmission coefficient (upper panel) and the zero bias current (lower panel), obtained from Eqs.~(\ref{a16}), (\ref{a17}) for $\epsilon=\pm 1$. We find that similar to symmetric double-dot, $\Delta T(E)$ changes its sign when $E$ is close to eigen-states of the double-dot. Moreover, the zero-bias current $I_{zb}(\mu )$ reaches its extremal values at the same values of the Fermi energy $\mu$, as for symmetric double-dot. In the same way, $I_{zb}(\mu )$ vanishes for $\mu\to\infty$.

\begin{figure}[h]
\includegraphics[width=8cm]{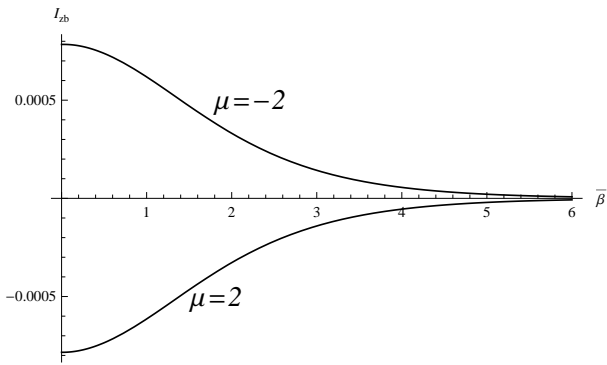}
\caption{(Color online) The zero bias current $I_{zb}$ as a function of the $U/\cT$ for two values of Fermi energy, $\mu=\pm 2$, corresponding to extremal points of the zero-bias current in Fig.~\ref{fig4}.  Other parameters are the same as in Fig.~\ref{fig4}. }
\label{fig6}
\end{figure}

Until now, our examples presented symmetric noise, corresponding to infinite noise-temperature, $\gamma_+=\gamma_-=\gamma/2$. Eq.~(\ref{relocc}). Figure~\ref{fig6} displays the zero bias current as a function of the inverse noise temperature $\bar\beta =U/\cT$.
One finds from this figure that as expected, the zero-bias current vanishes, when the noise-temperature, $\cT\to 0$, corresponding to $\gamma_-=0$. As a result, there are no transitions due to noise between the sub-levels, $\pm U$ of the left dot in Fig.~\ref{fig2}. As a result, $\Delta T(E)=0$.

\section{Interpretation}
\label{sec6}

It follows from our calculations that the zero-bias dc-current, generated by noise, would appear in double-dot coupled to Markovian leads. However, for the same conditions, the zero-bias current is {\em not} expected in single dot. What is the reason for such a different behavior of similar systems under the noise? A proper understanding of this point is very necessary. In particular, it can reveal physical origin of the zero-bias current produced by isotropic in space noise.

Consider first the resonant tunneling through a single quantum dot, Fig.~\ref{fg1}, coupled to Markovian leads, where the tunneling rates to the leads are very different, for instance $\Gamma_L\gg \Gamma_R$. Then the dwell-time (occupation probability) of electron inside the dot would be always much larger for an electron traveling from the left-to-right lead, than that from the right-to-left lead. Indeed, an electron from the left lead enters the dot very fast ($\sim 1/\Gamma_L$). Then it is trapped in the dot, since tunneling from the dot to the right lead is very slow ($\sim 1/\Gamma_R$). The situation is  opposite for the reversed case.

On first sight this looks as a violation of time-reversal symmetry in resonant tunneling. Yet, it is not the case, since penetration probability equals to occupation probability of the dot, multiplied by tunneling rate from the dot to the lead, Eq.~(\ref{transm4}). As a result, different dwell-times for direct and reversed processes are compensated by the subsequent tunneling rate to a corresponding lead. This restores the time-reversal symmetry of the transmission probabilities.

In the case of Markovian leads, the presence of noise does not violate the time-reversal symmetry of transmission probabilities, since the tunneling rates to the leads ($\Gamma_{L,R}$) are energy independent. As a result, fluctuations of the energy level ($E_0$) do not affect the tunneling rates, so that the time-reversal symmetry would hold.

Consider now the resonant tunneling through a double-dot system, Fig.~\ref{fig1}, starting from the no-noise case. Here too, the occupation probability of the left dot depends on whether the electron is coming from left or from right lead, even  if the double-dot is symmetric ($\Gamma_L=\Gamma_R$, $\epsilon =0$). Indeed, an electron from the left lead is coming directly to left dot, where an electron from the right lead have to cross the middle barrier. However, as in previous case,  such a difference in dwell-times is fully compensated by a subsequent tunneling  rate, restoring the time-reversal symmetry of transmission probabilities.

Now we include the noise, randomly fluctuating the  energy level of the left dot between two values, $\pm U/2$, Fig.~\ref{fig2}. These fluctuations cannot influence the energy-independent tunneling rates from the dots to leads, $\Gamma_{L,R}$. However, they  destroy the linear superposition of electron states between two dots. Note that any  quantum mechanical transitions between isolated states proceed through such a superposition of these states. The latter is described by the off-diagonal density-matrix element, $q_{12}^{(L,R)}$, Eq.~(\ref{qaver}). Since it depends on energy difference between the levels (see Eq.~(\ref{a15c})), the ensemble average of $q_{12}^{(L,R)}$ over the noise diminishes this quantity  (decoherence), unless the noise equally affects the both dots \cite{fn4}.

If the double-dot is isolated from leads, a single electron occupying such a  system  (qubit), is under the noise infinitely long time. Then its off-diagonal density matrix element vanishes due to decoherence, $q_{12}^{(L,R)}(t\to\infty)\to 0$, (c.f. with Ref.~[\onlinecite{amnon,gm}]. However, in the case of electron transport through the double-dot, each electron coming from the leads,  occupies the dots only by a finite time. Then the effect of decoherence would be  proportional to the electron dwell-time inside the dot, which is under the noise. Hence, the decoherence effect would be different for electrons, arriving the dot from the left lead than that for electrons arriving the dot from the right lead. This produces  violation of the time-reversal symmetry, resulting in zero bias current.

\begin{figure}[h]
\includegraphics[width=8cm]{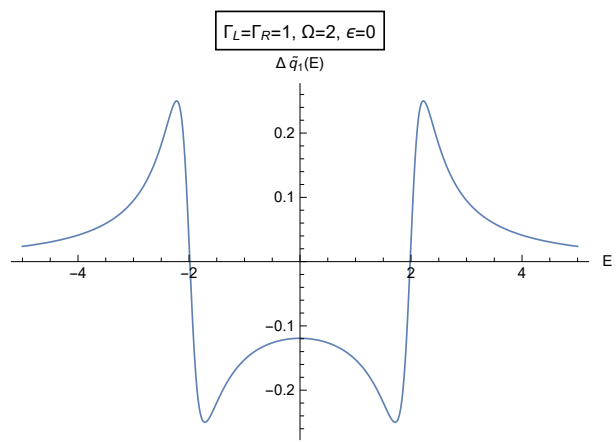}
\caption{(Color online) Difference in occupation of the first dot for an electron, coming from left or right lead, in the case of no-noise ($U=0$) }
\label{fig7}
\end{figure}

It follows from the above explanation that the difference in occupation of the left dot by an electron, coming from left and right lead, should be similar to the time-reversal violation of the transmission probability, $\Delta T(E)$, Eq.~(\ref{dT11}). Indeed, let us evaluate the difference in occupation of the first dot,
\begin{align}
\Delta {\tq}_1(E)=\Gamma_L{\tQ}_1^{(L)}(E)-\Gamma_R{\tQ}_1^{(R)}(E)
\end{align}
Since this quantity is not considerably affected by the noise, we do it
analytically for the case of no-noise, $U=0$. Using Eqs.~(\ref{BB1})-(\ref{BB3}), one easily obtains from for a symmetric double dot, $\Gamma_L=\Gamma_R=\Gamma/2$ and $\epsilon=0$, very simple analytical expressions
\begin{align}
\Delta {\tq}_1(E)={\Gamma\over2}\, {E^2+{\Gamma^2\over16}-\Omega^2\over
\big(E^2+{\Gamma^2\over16}-\Omega^2\big)^2
+{\Gamma^2\over4}\Omega^2}\, .
\label{BB4}
\end{align}
By confronting this expression with Eq.~(\ref{dT11}) we find that $\Delta T(E)$ and $\Delta {\tq}_1(E)$ are interrelated. It is illustrated in Fig.~\ref{fig7}.
Comparing this figure with Fig.~\ref{fig4} (upper panel), we find that the behaviour of $\Delta T(E)$ as a function of $E$ is very similar to that of $\Delta \tq_1(E)$. Altogether it confirms our understanding of the zero-bias current as the effect of decoherence, generated by an external noise.

\section {Discussion}
\label{sec7}

In this paper we investigated the zero-bias current, induced by a dichotomic (telegraph) noise, fluctuating energy levels of a quantum system. Our results were obtained by application of the generalized Landauer formula for the time-dependent non-interacting electron transport, assuming the absence of back-action from the quantum system to the noise. In this case the origin of noise becomes irrelevant, so that the results can be considered as applicable for any quantum system in the fluctuating environment.

We found that the necessary condition for the steady-state zero-bias dc-current through a quantum system is a break of the time-reversal symmetry in the transmission probability. If this symmetry holds, no zero-bias current is expected in such a system. We confirm it for the case of a single quantum dot coupled to Markovian leads, where the time-reversal symmetry persists, even in the presence of noise.

However, in the case of a double-dot, we found that the noise violates the time-reversal symmetry of the transmission coefficient, leading to zero-bias current. Our detailed analysis demonstrates that such a violation is a result of decoherence, generated by the noise. This always takes place when the current proceeds through linear superposition of isolated quantum states (delocalized orbitals). Since decoherence due to noise is ubiquitous phenomenon, we assume that similar zero-bias current can be found in many other quantum systems.

For instance, it can appear even in a single dot, coupled to {\em non-Markovian} leads of finite band-width. Indeed, we demonstrated in this paper that this case corresponds to a quantum dot, directly coupled to isolated (pseudo) modes, imbedded in a {\em Markovian} spectrum. As a result, the electron traveling through the dot, appears in a linear superposition of the dot and pseudo-mode states. This superposition is affected by decoherence due to noise, resulting in zero-bias current.

Our predictions can be verified experimentally by attaching a random voltage source to the system, for instance via a plunger. Alternatively, it can be done by coupling capacitively the quantum system at zero bias to an external fluctuator, represented by an impurity (quantum dot) in equilibrium with a heat bath\cite{13,alt,18}. The back action can be ignored when the fluctuator's dynamics is governed by its coupling to a thermalizing heat bath, which is much stronger than its coupling to the system. In an another (non-equilibrium) example, discussed ion Sec.~\ref{sec4}, a current flows between two reservoirs through a quantum dot, located near the system, Fig.~\ref{fg4}.

In this paper we did not consider the zero-bias electron current generating by a periodically oscillating energy level of quantum dot, for Markovian and non-Markovian leads \cite{liliana}. The most interesting question is whether it exists an essential difference in directed particle flow induced by a periodic ac-field versus random forces. This problem will be discussed in a separate work.

\appendix
\section{Shapiro-Loginov formula for asymmetric noise}\label{app1}

Since the use of Shapiro-Loginov differential formula\cite{shapiro} makes the SEA very effective tool for an account of the noisy environment, we present here an original (Shapiro and Loginov) derivation of this formula with an amendment, suited for asymmetric telegraph noise.

Consider a functional $R[\xi(t),t]$ of a random variable $\xi(t)$. The  average of this functional over all the possible trajectories $\{\xi(t)\}$ in a time-interval $(0,t)$ is denoted by $\langle R[\xi(t),t]\rangle$. By expanding $R[\xi(t),t]$ in a time-ordered functional Taylor series, we find \cite{shapiro}
\begin{align}
&R[\xi(t),t]=R[0,t]+\sum_{n=1}^\infty\int\limits_0^{t}
dt_1\int\limits_0^{t_1}dt_2\cdots
\int\limits_{0}^{t_{n-1}}dt_n\nonumber\\
&\times K_n(t,t_1,t_2,\ldots ,t_n)
\xi(t_1)\xi(t_2)
\cdots\xi(t_{n})\cdots\xi(t_{n})
\label{shl}
\end{align}
where
\begin{align}
K_n(t,t_1,t_2,\ldots ,t_n)={\delta^n R[\xi(t),t]\over \delta \xi_1\delta\xi_2\cdots\delta\xi_n}\Big|_{\xi\to 0}
\end{align}
$\xi_j=x(t_j)$ and $\delta/\delta\xi(t)$is a functional derivative.

Let us multiply Eq.~(\ref{shl}) by $\xi(t)$, represented the telegraph noise, Eqs.~(\ref{noise0})-(\ref{rate2}) and average it over all trajectories. One obtains
\begin{align}
&\langle\xi(t)R[\xi(t),t]\rangle=\bar\xi\, R[0,t]+
\sum_{n=1}^\infty\int\limits_0^{t}
dt_1\int\limits_0^{t_1}dt_2\cdots
\int\limits_{0}^{t_{n-1}}dt_n\nonumber\\
&\times K_n(t,t_1,t_2,\ldots ,t_n)
\lr\xi(t)\xi(t_1)\xi(t_2)
\cdots\xi(t_{n})\rr\, ,
\label{shl2}
\end{align}
where $\bar\xi=\lr\xi(t)\rr=(\gamma_-^{}-\gamma_+^{})/\gamma$, Eq.~(\ref{rate2}).

Consider the integrant of this expression. One can rewrite it explicitly as
\begin{align}
&\langle\xi(t)\xi(t_1)\cdots\xi(t_{n})\rangle =\sum_{\xi,\xi_1^{},\ldots,\xi_n^{}=\pm 1}\xi\xi_1^{}\xi_2^{}\cdots\xi_n^{}
P_{\xi,\xi_1^{}}(t,t_1)\nonumber\\
&\times P_{\xi_1^{},\xi_2^{}}(t_1,t_2)\cdots P_{\xi_{n-1}^{},\xi_{n}^{}}(t_{n-1},t_n)P_{\xi_n}(t_n)
\label{cond1}
\end{align}
where $P_{\xi_j^{},\xi_{j+1}^{}}(t_j,t_{j+1})$ denotes the conditional probability for finding $\xi_{j+1}$ at time $t_{j+1}$, where it was  $\xi_{j}$ at time $t_{j}$. Using Eq.~(\ref{rate1}) we can write
\begin{align}
P_{\xi,\xi_1}(t,t_1)&={\gamma_-^{}\delta_{\xi,1}
+\gamma_+^{}\delta_{\xi,-1}\over \gamma}\nonumber\\
&+\Big[\delta_{\xi,\xi_1}-{\gamma_-^{}\delta_{\xi,1}
+\gamma_+^{}\delta_{\xi,-1}\over \gamma}\Big]e^{-\gamma (t-t_1)}
\label{cond2}
\end{align}
Therefore
\begin{align}
&\sum_{\xi=\pm 1} \xi P_{\xi,\xi_1}(t,t_1)=
{\gamma_-^{}-\gamma_+^{}\over \gamma}\nonumber\\
&+\sum_{\xi=\pm 1} \xi\Big[\delta_{\xi,\xi_1}-{\gamma_-^{}\delta_{\xi,1}
+\gamma_+^{}\delta_{\xi,-1}\over \gamma}\Big]e^{-\gamma (t-t_1)}
\end{align}

Differentiating this expression by time $t$ and substituting this result into Eq.~(\ref{cond1}) we find
\begin{align}
&{d\over dt}\langle\xi(t)\xi(t_1)\cdots\xi(t_{n})\rangle =-\gamma \langle\xi(t)\xi(t_1)\cdots\xi(t_{n})\rangle\nonumber\\
&~~~~~~~~~~~~~~~~~~~~~~~
+(\gamma_-^{}-\gamma_+^{})
\langle\xi(t_1)\cdots\xi(t_{n})\rangle
\end{align}
Using this result and Eq.~(\ref{shl2}) (see also in Appendix of Ref.~[\onlinecite{GAE}]), we easily arrive to Eq.~(\ref{df}).

\section{Derivation of single-electron Master equations}\label{app2}

Consider Eqs.~(\ref{a66}) for the amplitudes $b_{1(2)}^{(\alpha)}(t)$. Multiplying these equation on $b_{1(2)}^{(\alpha)*}(t)$ and subtracting the complex conjugated equations, one finds
\begin{subequations}
\label{a11}
\begin{align}
&{d\over dt} q_1^{(\alpha)}(E,t)=-\Gamma_L\, q_1^{(\alpha)}-2\Omega\,{\rm Im}[q_{12}^{(\alpha)}]
\nonumber\\
&~~~~~~~~~~~~~~~~~~~~~~~~~~~~~~~~~~~~
-2\,\Omega_\alpha{\rm Im}[ b_{1}^{(\alpha)}]\,\delta_{\alpha L}
\label{a10a}\\
&{d\over dt}\, q_2^{(\alpha)}(E,t) =-\Gamma_R\,  q_2^{(\alpha)}+2\Omega\,{\rm Im}[ q_{12}^{(\alpha)}]
\nonumber\\
&~~~~~~~~~~~~~~~~~~~~~~~~~~~~~~~~~~~~
-2\,\Omega_\alpha{\rm Im}[ b_{2}^{(\alpha)}]\,\delta_{\alpha R}\label{a11b}\\
&{d\over dt}\, q_{12}^{(\alpha)}(E,t)=i\Big[E_2(t)-E_1(t)
+i{\Gamma\over2}\Big] q_{12}^{(\alpha)}\nonumber\\
&~~~~~
+i\Omega\big[ q_1^{(\alpha)}- q_2^{(\alpha)}\big]
-i\Omega_\alpha \big[b_2^{(\alpha)*}\delta_{\alpha L}- b_1^{(\alpha)}\delta_{\alpha R}\big]
\label{a11c}
\end{align}
\end{subequations}
where $\Gamma=\Gamma_L+\Gamma_R$ and  $q_{1(2)}^{(\alpha)}\equiv q_{1(2)}^{(\alpha)}(E,t)=|b_{1(2)}^{(\alpha)}(E,t)|^2$,  $q_{12}^{(\alpha)}\equiv q_{12}^{(\alpha)}(E,t)=
b_{1}^{(\alpha)}(E,t)b_{2}^{(\alpha)*}(E,t)$. In the case of fluctuating level of the left dot (Fig.~\ref{fig2}), $E_1(t)=\pm U/2$ and $E_2(t)=\epsilon$, Eqs.~(\ref{a11}), averaged over the noise, read
\begin{subequations}
\label{a11p}
\begin{align}
&{d\over dt}\,\lr q_1^{(\alpha)}(E,t)\rr=-\Gamma_L\, \lr q_1^{(\alpha)}\rr-2\Omega\,{\rm Im}[\lr q_{12}^{(\alpha)}\rr ]\nonumber\\
&~~~~~~~~~~~~~~~~~~~~~~~~~~~~~~~~~~~
-2\,\Omega_\alpha{\rm Im}[\lr b_{1}^{(\alpha)}\rr]\,\delta_{\alpha L}
\label{a11pa}\\
&{d\over dt}\, \lr q_2^{(\alpha)}(E,t)\rr =-\Gamma_R\, \lr q_2^{(\alpha)}\rr+2\Omega\,{\rm Im}[\lr q_{12}^{(\alpha)}\rr]\nonumber\\
&~~~~~~~~~~~~~~~~~~~~~~~~~~~~~~~~~~~
-2\,\Omega_\alpha{\rm Im}[\lr b_{2}^{(\alpha)}\rr]\,\delta_{\alpha R}\label{a11pb}\\
&{d\over dt}\,\lr q_{12}^{(\alpha)}(E,t)\rr=i\Big[\epsilon
+i{\Gamma\over2}\Big]\lr q_{12}^{(\alpha)}\rr
-i{U\over2}\lr q_{12\xi}^{(\alpha)}\rr\
\nonumber\\
&
+i\Omega\big[\lr q_1^{(\alpha)}\rr-\lr q_2^{(\alpha)})\rr\big]
-i\Omega_\alpha \big[\lr b_2^{(\alpha)*}\rr\delta_{\alpha L}-\lr b_1^{(\alpha)}\rr\delta_{\alpha R}\big]
\label{a11pc}
\end{align}
\end{subequations}
where $\lr q_{12\xi}^{(\alpha)}\rr\equiv \lr\xi(t)q_{12}^{(\alpha)}(E,t)\rr$. To evaluate this term, we use the same procedure as in Eqs.~(\ref{a9}), (\ref{a10}), together with the Shapiro-Loginov differential formula, Eq.~(\ref{df}), thus obtaining
\begin{subequations}
\label{a12}
\begin{align}
&{d\over dt}\,\lr q_{1\xi}^{(\alpha)}(E,t)\rr=-(\Gamma_L+\gamma)\, \lr q_{1\xi}^{(\alpha)}\rr+\gamma\bar\xi \lr q_{1}^{(\alpha)}\rr \nonumber\\
&~~~~~~~~~~~~~~
-2\Omega\,{\rm Im}[\lr q_{12\xi}^{(\alpha)}\rr ]-2\,\Omega_\alpha{\rm Im}[\lr b_{1\xi}^{(\alpha)}\rr]\,\delta_{\alpha L}\label{a12a}\\
&{d\over dt}\, \lr q_{2\xi}^{(\alpha)}(E,t)\rr =-(\Gamma_R+\gamma)\, \lr q_{2\xi}^{(\alpha)}\rr+\gamma\bar\xi \lr q_{2}^{(\alpha)}\rr \nonumber\\
&~~~~~~~~~~~~
+2\Omega\,{\rm Im}[\lr q_{12\xi}^{(\alpha)}\rr]-2\,\Omega_\alpha{\rm Im}[\lr b_{2\xi}^{(\alpha)}\rr]\,\delta_{\alpha R}\label{a12b}\\
&{d\over dt}\,\lr q_{12\xi}^{(\alpha)}\rr(E,t)=i\Big[\epsilon
+i{\Gamma+2\gamma\over2}\Big]\lr q_{12\xi}^{(\alpha)}\rr
+\Big(\gamma\bar\xi-i{U\over2}\Big) \lr q_{12}^{(\alpha)}\rr\nonumber\\
&
+i\Omega\big[\lr q_{1\xi}^{(\alpha)}\rr-\lr q_{2\xi}^{(\alpha)}\rr\big]
-i\Omega_\alpha\big[\lr b_{2\xi}^{(\alpha)*}\rr\delta_{\alpha L}-\lr b_{1\xi}^{(\alpha)}\rr\delta_{\alpha R}\big]
\label{a12c}
\end{align}
\end{subequations}
where $\lr q_{1\xi,2\xi}^{(\alpha)}(E,t)\rr =\lr\xi(t)q_{1,2}^{(\alpha)}(E,t)\rr$ and $\lr q_{12\xi}^{(\alpha)}(E,t)\rr =\lr\xi(t)q_{12}^{(\alpha)}(E,t)\rr$.

Now we introduce new variables (c.f. with Eq.~(\ref{a13a}))
\begin{align}
&q_{1\pm(2\pm)}^{(\alpha)}(E,t)=\lr q_{1(2)}^{(\alpha)}(E,t)\rr\pm  \lr q_{1\xi(2\xi)}^{(\alpha)}(E,t)\rr\nonumber\\
&q_{12\pm}^{(\alpha)}(E.t)=\lr q_{12}^{(\alpha)}(E,t)\rr\pm \lr q_{12\xi}^{(\alpha)}(E,t)\rr
\label{a13b}
\end{align}
In these variables, Eqs.~(\ref{a12}) turn to  Eqs.~(\ref{a15}), representing the Master equations for  single-electron transport through a double-dot system.

\begin{acknowledgments}
I thank Amnon Aharony, Ora Entin-Wohlman, Xin-Qi Li, Rafael S\'{a}nchez and Robert Shekhter for useful discussions.
\end{acknowledgments}


\begin{thebibliography}{999}
\bibitem{fn0} We use the term ``Markovian'' in a relation to the spectral-density function, although in the literature it can be related to density of state only.
\bibitem{fn5} Our calculations of the energy current within the SEA (not presented in this paper), are in full agreement with the results of Ref.~[\onlinecite{ora}].
\bibitem{fn2} We do not consider the initial state, corresponding to occupied double-dot, since it does not contribute to steady-state current.
\bibitem{fn4}
    Similar mechanism of decoherence is discussed in \cite{amnon}, for a single qubit, imbedded in the AB ring.
\end{thebibliography}
\end{document}